\documentclass[12pt]{iopart}
\usepackage[utf8]{inputenc}

\usepackage{iopams}  

\expandafter\let\csname equation*\endcsname\relax

\expandafter\let\csname endequation*\endcsname\relax

\usepackage{graphicx,epsfig,amssymb,times} 
\usepackage{amsmath,amsfonts}
\usepackage{bm}
\usepackage{epstopdf}
\usepackage{hyperref}
\usepackage[caption=false]{subfig}
\usepackage[usenames]{color}   
\usepackage[dvipsnames]{xcolor}
\usepackage{multirow}
\usepackage[normalem]{ulem}

\newcommand{\dd}{\text{d}}
\newcommand{\ii}{\text{i}}

\definecolor{coolblack}{rgb}{0.0, 0.18, 0.39}
\definecolor{darkred}{rgb}{0.5,0,0}
\definecolor{darkgreen}{rgb}{0,0.5,0}
\definecolor{darkblue}{rgb}{0,0,0.5}
\definecolor{lapislazuli}{rgb}{0.15, 0.38, 0.61}
\definecolor{venetianred}{rgb}{0.78, 0.03, 0.08}
\definecolor{bleudefrance}{rgb}{0.19, 0.55, 0.91}
\definecolor{dogwoodrose}{rgb}{0.84, 0.09, 0.41}
\hypersetup{colorlinks=true, citecolor=darkblue, linkcolor=darkblue, 
urlcolor = darkblue}

\begin{document}
\title[Gravitational and kinematic redshifts of black holes in nonlinear electrodynamics]{Gravitational and kinematic redshifts of black holes in nonlinear electrodynamics}

\author{Marco A. A. de Paula$^{\color{blue}1,2}$ and Mustapha Azreg-A\"{\i}nou$^{\color{blue}3}$}

\address{$1$\ Faculdade de Ciências Naturais, Universidade Federal do Par\'a, Campus Universit\'ario do Tocantins-Cametá, 68400-000, Camet\'a, Par\'a, Brazil}

\address{$2$\ Departamento de F\'isica, Universidade Federal da Para\'iba, 58051-970, Jo\~ao Pessoa, Para\'iba, Brazil.}

\address{$3$\ Ba\c{s}kent University, Engineering Faculty, Ba\u{g}l{\i}ca Campus, 06780-Ankara, T\"{u}rkiye}

\eads{\color{blue}\mailto{marcodepaula@ufpa.br}, \mailto{azreg@baskent.edu.tr}}

\begin{abstract}
	
Spacetimes arising from nonlinear electrodynamics (NED) are a good laboratory for studying both the nature of regular black hole (RBH) solutions and the imprints of nonlinear electromagnetic fields within this context. Over the past few decades, NED-sourced black hole (BH) spacetimes have attracted considerable attention, but electrically charged RBHs obtained in the so-called $F$ framework have been less addressed in the literature. We consider two electrically charged, fully regular members of the h-family that are solutions to general relativity minimally coupled to NED. Because of their potential astrophysical and astronomical applications, we mainly focus our investigation on the motion of photons and its implications for the gravitational and kinematic redshifts, as well as for the shadow radius, considering the effective geometry followed by photons in NED. Moreover, we consider the observational data for Sagittarius A$^{\star}$ and Messier 87$^{\star}$ BHs to impose some constraints on the charge-to-mass ratio of the fully RBHs. We obtain the general equations that govern the gravitational and kinematic redshifts of NED-sourced BH spacetimes. In particular, for fully RBHs with BH charge-to-mass ratios above some moderate value, we observe discrepancies in their physical and geometrical properties compared to those of the Reissner--Nordstr\"om (RN) BH. Therefore, our results indicate that it is possible to distinguish NED-based RBH solutions from RN BHs, from the perspective of gravitational and kinematic redshifts, suggesting potential astrophysical relevance.
	
	\vspace{0.5cm}
	\vspace{0.5cm}\\(Some figures may appear in colour only in the online journal)
	
\end{abstract}

%\submitto{\CQG}

%%%%%%%%%%%%%%%%%%%%%%%%%%%%%%%%%%%%%%%%%%%%%%%%%%
\section{Introduction}
%%%%%%%%%%%%%%%%%%%%%%%%%%%%%%%%%%%%%%%%%%%%%%%%%%

Singularities in a spatial region are the places where the laws of nature break down. Such an effect is not welcomed in physics and, in general, provides an indication that the physical theory behind it has drawbacks. Singularities emerge in the theory of general relativity (GR), as they do in linear electrodynamics (LED). For example, in GR, singularities may appear as pathologies in spacetime when curvature scalars blow up~\cite{wald2010general}, while in LED, the self-energy of a point charge can be unbounded~\cite{feynman2011feynman}. The coupling of nonlinear GR with LED has not resulted in the removal of metric and LED field singularities, even when other normal or phantom fields (scalars in general) are considered, as in Einstein-Maxwell-dilaton theory~\cite{Fabris2009}. In the late 1990s, Eloy Ay\'{o}n-Beato and Alberto Garc\'{\i}a realized that it was possible to obtain exact charged regular black hole (RBH) solutions, i.e., black holes (BHs) free of curvature singularities, based on the minimal coupling between GR and models of nonlinear electrodynamics (NED)~\cite{Ayon-Beato:1998hmi}. This theory was originally perceived as an attempt to generalize Maxwell's theory in the ultraviolet regime~\cite{born1934foundations, born1934quantum}. However, nowadays, it is commonly applied in BH physics as a mathematical tool or mechanism that allows the derivation of RBH solutions. Many investigations have gone in the direction of showing how this mechanism actually works~\cite{Bronnikov:2000vy,Bronnikov:2024}.

NED Lagrangians may either depend explicitly on the electric and/or magnetic charge, as do the Ay\'{o}n-Beato and Garc\'{\i}a Lagrangian~\cite{Ayon-Beato:1998hmi} and the Lagrangian that produces the source for the Bardeen RBH~\cite{bardeen1968non,Ayon-Beato:2000mjt}, or be independent of the charge~\cite{Sorokin:2021tge,Bronnikov:2024b}, as does the Born-Infeld Lagrangian~\cite{born1934foundations}. Moreover, NED models can be represented in two related frameworks: the so-called $F$ and $P$ frameworks~\cite{Bronnikov:2000vy}. In the $F$ framework, the field system is expressed in terms of a gauge-invariant Lagrangian $\mathcal{L}(F)$, where $F=F_{\mu\nu}F^{\mu\nu}$ is the Maxwell scalar, with $F_{\mu\nu}$ being the standard electromagnetic field tensor. The dual formalism allows us to convert the $\mathcal{L}(F)$ Lagrangian to an $\mathcal{L}(P)$ one by means of a Legendre transformation~\cite{Gutierrez:1981ed}, leading to the $P$ framework, where $P = P_{\mu\nu}P^{\mu\nu}$, and $P_{\mu\nu}$ is an auxiliary electromagnetic field tensor, given by $P_{\mu\nu}=\mathcal{L}_F F_{\mu\nu}$~\cite{Balart:2014plb}. The $P$ framework is commonly used to obtain electrically charged RBHs.

Incorporation of NED stress-energy tensors as sources into the field equations has, in addition to smoothing the singularities, the effect of modifying the motion of photons~\cite{Novello:1999pg}. The paths of photons are geodesics of some effective geometry (EG), $\bar{g}_{\mu\nu}$, which depends on the NED fields and is related to the background geometry, $g_{\mu\nu}$, by $\bar{g}_{\mu\nu}=\alpha_{\mu\nu}g_{\mu\nu}$ (no summation), where $\alpha_{\mu\nu}$ are functions of the NED fields~\cite{dePaula:2023ozi}. Consequently, all that is related to the motion of photons in NED must take the EG into account to achieve the correct results. Over the past few years, several works have investigated the role driven by photons in NED-sourced spacetimes in the context of gravitational lensing~\cite{dePaula:2023ozi,Stuchlik:2019uvf,Habibina:2020msd,Toshmatov:2021fgm,Allahyari:2019jqz,Hu:2020usx,Eiroa:2005ag} and thermodynamics~\cite{Balart:2017,Abe:2025vdj,Aydiner:2025eii,Bokulic:2021dtz,Barrientos:2025rde,Hale:2025veb}. Moreover, the causality condition is the requirement that the trajectories of photons are timelike in the background metric $g_{\mu\nu}$. This results in inequality constraints and theorems concerning the NED fields~\cite{Schellstede:2016zue,Russo:2024llm,dePaula:2024yzy}.

In recent years, certain classic tests of Einstein's theory have been widely used to better understand different alternative distributions of matter in GR and beyond~\cite{Junior:2023nku,Bhattacharya:2025qcd,Hsieh:2021rru,Ben-salem:2022txj,Hook:2017psm,Poddar:2021sbc,Hou:2017cjy,DellaMonica:2023ydm,Fathi:2019jgd,Ashtekar:1975wt,Souza:2024ltj,Moradpour:2026ppq,Aliberti:2024udm}, in particular considering the Sagnac effect~\cite{sagnac1913ether,sagnac1913preuve} and the Shapiro time delay~\cite{Shapiro:1964uw,Shapiro:1968zza}. Another interesting classical test of GR --- yet less explored in the context of NED-sourced spacetimes --- is the gravitational and kinematic redshifts. Roughly speaking, in general, a star moving in a circular orbit emits radiation that can be measured by a detector, and the redshift depends on the 4-momentum of the photon emitted by the star and received by the detector~\cite{hobson2006general}. In this context, one can use the gravitational redshift to better understand the role played by the parameters describing the compact object and the theories beyond it (see, e.g., Ref.~\cite{martinez2024observational} and references therein). For example, by comparing the gravitational and kinematic redshift results between regular and singular BHs, it is possible to establish a clear way to distinguish between them.

For NED-sourced spacetimes, the study of gravitational and kinematic redshifts takes on an even more interesting contrast. First, since photons in NED propagate along null geodesics of an EG, the gravitational and kinematic redshifts are expected to be affected in this scenario, as discussed, e.g., in Refs.~\cite{MosqueraCuesta:2004em, Guzman-Herrera:2023zsv}~\footnote{These topics have also been drawing attention in alternative theories of gravity, see, e.g., Ref.~\cite{AraujoFilho:2025hnf} and references therein, but considering only the background geometry associated with NED-sourced spacetimes.}. Second, the gravitational and kinematic redshifts are particularly significant in the environments of neutron stars, especially magnetars. Vacuum birefringence, a phenomenon associated with NED models, is expected to be detectable in the vicinity of such objects due to the strength of their magnetic field~\cite{Mignani:2016fwz,Capparelli:2017mlv,STAR:2019wlg}. Third, there are charged compact stars that can harbor a huge amount of charge ($\sim 10^{20}C$)~\cite{Ray:2003gt} associated with very strong electric fields~\cite{gurtug2020gravitational}. Therefore, taking the effective geometry into account helps us uncover signatures of NED fields in BH physics. In addition to that, an analysis such as the one presented here provides important first steps toward a better understanding of the astrophysical scenarios in which NED fields may be potentially relevant.

Notice also that, although astrophysical BHs are typically expected to have significant angular momentum but negligible charge \cite{Gibbons:1975kk}, the role played by the charge cannot be simply overlooked in the astrophysical debate. For example, BHs formed by the collapse of a charged compact star can acquire primordial charge~\cite{Ray:2003gt}, and it is not clear on which timescales such a charged BH discharges. In addition to that, there are several mechanisms that can load or induce electric charges~\cite{Zajacek:2018ycb}, such as the rotation of a BH in an external magnetic field, the twisting of magnetic field lines threading the event horizon of rotating BHs, and accretion of plasma. Moreover, astrophysical BHs cannot be simply treated as nearly uncharged, although the corresponding BH charge is small, considering astrophysical BH systems~\cite{Zajacek:2018vsj,Levin:2018mzg}, and even a tiny amount of charge can significantly affect the dynamics of charged particles in the BH vicinity~\cite{Schroven:2017jsp}. Consequently, the role of charge in astrophysical scenarios is not entirely negligible, and investigations, even in spherically symmetric setups, are important for a better understanding of this issue.

The static spherically symmetric (SSS) solution derived in~\cite{Ayon-Beato:1998hmi}, considering the $P$ framework, has a finite metric (the metric components are non-singular and well-behaved) and NED fields for all values of the radial coordinate. In contrast, most known electrically charged solutions in NED minimally coupled to gravity, obtained in the $F$ framework, have a divergent electric field at the center $r=0$~\cite{Bronnikov:2022ofk}. Recently, in addition to mass and electric charge, a one-parameter family of SSS and fully RBHs (h-family) with $-2<h\leq 2$, where both the metric function and the electric field are regular for all values of the radial coordinate, has been determined~\cite{Azreg-Ainou:2025tuj}. Moreover, the fully RBH solutions of the h-family satisfy the weak energy condition. These spacetimes provide a good test ground for investigating the properties of electrically charged RBHs obtained in the $F$ framework, as well as NED fields in this context.  Here, our aim is to address the gravitational and kinematic redshifts of two members of the h-family, the $h=0$ and $h=2$ members~\footnote{Since the derivations of the RBH solutions for $h = 0$ and $h = 2$ have already been presented in Ref.~\cite{Azreg-Ainou:2025tuj}, we will consider them here, but we could consider other cases, e.g., $h = 1$, without further complications.}, considering the EG. We highlight that this class of solutions circumvents the no-go theorem for NED-based RBH solutions obtained in Ref.~\cite{Bronnikov:2000vy}. In ~\ref{appx}, we discuss this in detail.

To achieve this aim, we will conduct the following investigations. In Sec.~\ref{secpre}, we introduce some preliminaries and define our working ansatzes. In Sec.~\ref{sec:background}, we review the physical and geometric properties of the fully regular solution corresponding to $h=0$ and $h = 2$. We elaborate on the motion of photons, considering the EG, in Sec.~\ref{sec:mp}, studying the geodesic equations in ~\ref{sech2mp} and the weak deflection angle in~\ref{sech2wd}. We also use observational data for Sagittarius A$^{\star}$ (Sgr A$^{\star}$) and Messier 87$^{\star}$ (M87$^{\star}$) BHs to impose some astrophysical constraints on the BH charge-to-mass ratio of fully RBHs in~\ref{sech2ac}. Then, because of the potential relevance of NED fields in astronomy, we investigate the gravitational and kinematic redshifts in Sec.~\ref{sech2red}. We draw our conclusions in Sec.~\ref{sec:remarks}. Throughout this work, we take $c= 8\pi G=1$ and use the metric signature $(+, -, -, -)$, unless otherwise stated.

%%%%%%%%%%%%%%%%%%%%%%%%%%%%%%%%%%%%%%%%%%%%%%%%%%
\section{Preliminaries}\label{secpre}
%%%%%%%%%%%%%%%%%%%%%%%%%%%%%%%%%%%%%%%%%%%%%%%%%%

In the $F$ framework, the field equations are derived by varying the action~\cite{Bronnikov:2000vy}
\begin{equation}
	\label{1}S=\frac{1}{2}\int \dd ^4x\sqrt{|g|}~\left[R-\mathcal{L}(F)\right]\,,
\end{equation}
where $g$ is the determinant of the metric tensor \textbf{g}, $R$ is the Ricci scalar, and $\mathcal{L}(F)$ represents the NED Lagrangian. For convenience, we assume a general SSS metric of the form
\begin{equation}
	\label{sss0}\dd s^2=A(r)\dd t^2 + B(r) \dd r^2 + C(r)\dd \Omega^2\,,
\end{equation}
in which $A(r)$, $B(r)$, and $C(r)$ are metric components to be determined by the field equations, and $\dd \Omega^2 = \dd \theta^{2}+ \sin^{2}\theta \dd \varphi^{2}$ is the line element of a 2-sphere of unit radius.

The line element of an SSS metric associated with NED-sourced BH spacetimes can be written as
\begin{align}
	\label{2}\dd s^{2} & \equiv g_{\mu\nu}\dd x^{\mu}\dd x^{\nu} =f(r)\dd t^2 - f(r)^{-1}\dd r^2 - r^2\dd \Omega^2\,,    
\end{align}
for which $A(r) = - B(r)^{-1} = f(r)$, with $f(r)$ being the metric function, $C(r) = -r^{2}$, and the field equations lead to
\begin{align}
	\label{3a}&\frac{rf(r)^{\prime}+f(r)-1}{r^2}=-\frac{\mathcal{L}(F)}{2}+F\mathcal{L}_F\,,\\
	\label{3b}&\frac{rf(r)^{\prime\prime}+2f(r)^{\prime}}{2r}=-\frac{\mathcal{L}(F)}{2}\,,
\end{align}
and
\begin{equation}
	\label{3c}\nabla_{\mu}(\mathcal{L}_F~F^{\mu\nu})=0\,,
\end{equation}
where $\mathcal{L}_F=\dd \mathcal{L}/\dd F$ and the prime notation denotes the derivative with respect to the radial coordinate $r$. Note that Eq.~\eqref{3c} is a conservation equation for the Faraday tensor, and this tensor also satisfies the Bianchi identity. 

Since in this work we are interested in purely electrically charged solutions, Eq.~\eqref{3c} is easily integrated and yields the electric field expression $E(r)$, given by
\begin{equation}
	\label{electric field}E(r) = F^{tr}=\frac{Q}{r^2\mathcal{L}_F}\,,
\end{equation}
along with that of the invariant $F$, namely
\begin{equation}
	\label{invariantF}F=-2(F^{tr})^2=-\frac{2Q^2}{r^4\mathcal{L}_F^2}\,, 
\end{equation}
with $Q$ denoting the electric charge (see, e.g., Ref.~\cite{Bokulic:2025usc} for the definition of electric and magnetic charges in NED-sourced spacetimes). In the $F$ framework, most known electrically charged solutions in NED minimally coupled to gravity have a divergent electric field at the center $r=0$~\cite{Bronnikov:2022ofk}.  This singularity can be removed at the price of choosing $\mathcal{L}_F$ divergent at the center, but we can preserve $\mathcal{L}(F)$. There are many ways to remove the singularity at the center~\cite{Azreg-Ainou:2025tuj}. In this work, we choose $\mathcal{L}_F$ in the form
\begin{equation}
	\label{4b}\mathcal{L}_F=\frac{1}{x^8}+\frac{h}{x^4}+1\,,
\end{equation}
where $h$ is a dimensionless real parameter and $x$ is a dimensionless radial coordinate defined in terms of $r$ by
\begin{equation}
	\label{x}x=\frac{r}{K}\,,	
\end{equation}
in which $K = K(M,Q)$ is a positive parameter that has the dimension of length and will be given in terms of the ratio $Q^2/M$ in the subsequent sections, with $M$ being the Arnowitt-Deser-Misner (ADM) mass (see, e.g., Refs.~\cite{Fan:2016hvf,Toshmatov:2018cks} for details). For the sake of simplicity, we have chosen to present $\mathcal{L}(F)$ and its derivatives with respect to $F$ in a parametric form in terms of $x$. This is due to the complexity of Eq.~\eqref{invariantF} in this context, which results in a lengthy equation for $\mathcal{L}(F)$. However, expressions in terms of $F$ are presented in Ref.~\cite{Azreg-Ainou:2025tuj}. Here, we focus on the more phenomenological aspects of the model. The expressions given above of $F$~\eqref{invariantF} and $\mathcal{L}_F$~\eqref{4b} yield an expression for the product $F\mathcal{L}_{FF}$ (needed for subsequent sections), where $\mathcal{L}_{FF}=\partial_x \mathcal{L}_{F}/\partial_x F$, i.e.,
\begin{equation}
	\label{FLFF}F\mathcal{L}_{FF}=\frac{(h x^4+2) (x^8+h x^4+1)}{x^8 (x^8-h x^4-3)}\,.
\end{equation}

As noted in~\cite{Azreg-Ainou:2025tuj}, the electric charge of the solution is not local; rather, it is distributed spatially with a non-delta distribution function, in contrast to the Reissner--Nordstr\"om (RN) solution, where the electric charge has a delta distribution. To show that, apply the Gauss theorem to $E(r)$~\eqref{electric field} and obtain the electric charge $\mathcal{Q}(r)$ inside the sphere of radius $r=Kx$, \[\mathcal{Q}(r)=\frac{Q}{\mathcal{L}_{F}(r)}\,,\] and this yields an electric charge density \[\rho_q(r)=\frac{\partial_r \mathcal{Q}(r)}{4\pi r^2}=\frac{Q x^5 (2+h x^4)}{\pi  K^3 (1+h x^4+x^8)^2},\] which is finite for all $x$. The total charge of the distribution is $\int_{0}^{\infty}\rho_q(r)4\pi r^2\dd r=\mathcal{Q}(\infty)-\mathcal{Q}(0)=Q$. It is worth mentioning that $\rho_q(r)$ vanishes at both limits, i.e., $r\to 0$ and $r\to \infty$, so the electric charge density must have a finite maximum value at some $r_{\text{max}}(h)$ (which can be determined analytically as a function of $h$). This behavior of $\rho_q(r)$ is qualitatively different from its corresponding RN charge density, which is a Dirac-delta distribution at $r=0$.

Note that in the Maxwellian limit, we have $\mathcal{L}\to F$ (instead of $\mathcal{L}\to -F$), so that, in our case, the causality requirement that the trajectories of photons are timelike in the background metric $g_{\mu\nu}$ takes the form~\cite{Sorokin:2021tge}
\begin{equation}\label{CR1}
	\frac{\partial^2 \mathcal{L}}{\partial E^2}\leq 0\,,	
\end{equation}
where $E$ is the electric field defined in Eq.~\eqref{electric field}. Using $F=-2E^2$ and $\partial \mathcal{L}/\partial E = \mathcal{L}_F (\partial F/\partial E)$, we obtain
\begin{align}
	\label{CR2}\frac{\partial^2 \mathcal{L}}{\partial E^2}= -4( \mathcal{L}_F + 2F\mathcal{L}_{FF}).
\end{align}
Considering Eq.~\eqref{4b}, we get
\begin{equation}
	\frac{\partial^2 \mathcal{L}}{\partial E^2}= -\frac{4 (x^8+h x^4+1)^2}{x^8 (x^8-h x^4-3)}\,,
\end{equation}
which is manifestly negative if
\begin{equation}
	\label{CR3}r>r_\text{CR}\equiv K[(h+\sqrt{12+h^2})/2]^{1/4}\,,	
\end{equation}
where CR means causality requirement. Note that $r_\text{CR}$ is generally smaller than the radius of the outer horizon.

%%%%%%%%%%%%%%%%%%%%%%%%%%%%%%%%%%%%%%%%%%%%%%%%%%
\section{Backgrounds}\label{sec:background}
%%%%%%%%%%%%%%%%%%%%%%%%%%%%%%%%%%%%%%%%%%%%%%%%%%

In this section, we present the spacetimes that will be explored in this work, namely $h=0$ and $h=2$ of the h-family (see the Appendix of~\cite{Azreg-Ainou:2025tuj}), discussing mainly their metric functions and KS. For simplicity, we define the normalized electric charge, given by
\begin{equation}
	\label{nc}q \equiv \dfrac{Q}{Q_{\text{ext}}},
\end{equation}
which satisfies $0 \leq q \leq 1$ for BH solutions, where $Q_{\text{ext}}$ is the extremal charge value of the corresponding BH solution.

%%%%%%%%%%%%%%%%%%%%%%%%%%%%%%%%%%%%%%%%%%%%%%%%%%
\subsection{Fully RBH spacetimes with h = 0}\label{sech0}
%%%%%%%%%%%%%%%%%%%%%%%%%%%%%%%%%%%%%%%%%%%%%%%%%%

For $h=0$, the NED Lagrangian that produces an asymptotically flat metric is
\begin{equation}
	\label{n83}\mathcal{L}(x)=\dfrac{5184 M^4}{\pi^4 (\sqrt{2}+1)^2Q^6}\left(\dfrac{\pi}{2}-\frac{2 x^4}{ 1+x^8}- \arctan  x^4\right)\,,
\end{equation}
where $x$ is defined in~\eqref{x}, with the parameter $K$ given by
\begin{equation}
	\label{h0K}K:=\frac{\sqrt{\sqrt{2}+1} ~\pi Q^2}{6\times 2^{1/4} M}\,.
\end{equation}
The corresponding metric function is given by~\cite{Azreg-Ainou:2025tuj}
\begin{equation}
	\label{n84a}f(x)=1-\frac{36 \sqrt{2} (\sqrt{2}-1) M^2}{\pi^2 Q^2}~p(x)\,,	
\end{equation}
with
\begin{align}
	\nonumber p(x)= & \ \frac{1}{6 x}\Big\{ x^3 (\pi - 2\arctan  x^4) -\sum _{i=0}^3 \cos (7\alpha_i) \ln \Big[1+x^2-2(\cos
	\alpha_i) x \Big]\\
	\label{n84b}&+2 \sum _{i=0}^3 \sin (7\alpha_i) \arctan \Big[(\csc \alpha_i) x-\cot \alpha_i\Big]\Big\}\,,
\end{align}
and
\begin{equation}
	\alpha_i=\frac{(2 i+1) \pi }{8},
\end{equation}
where $i$ is an integer number that ranges from 0 to 3.

\begin{figure}[!htbp]
	\begin{centering}
		\includegraphics[width=0.6\columnwidth]{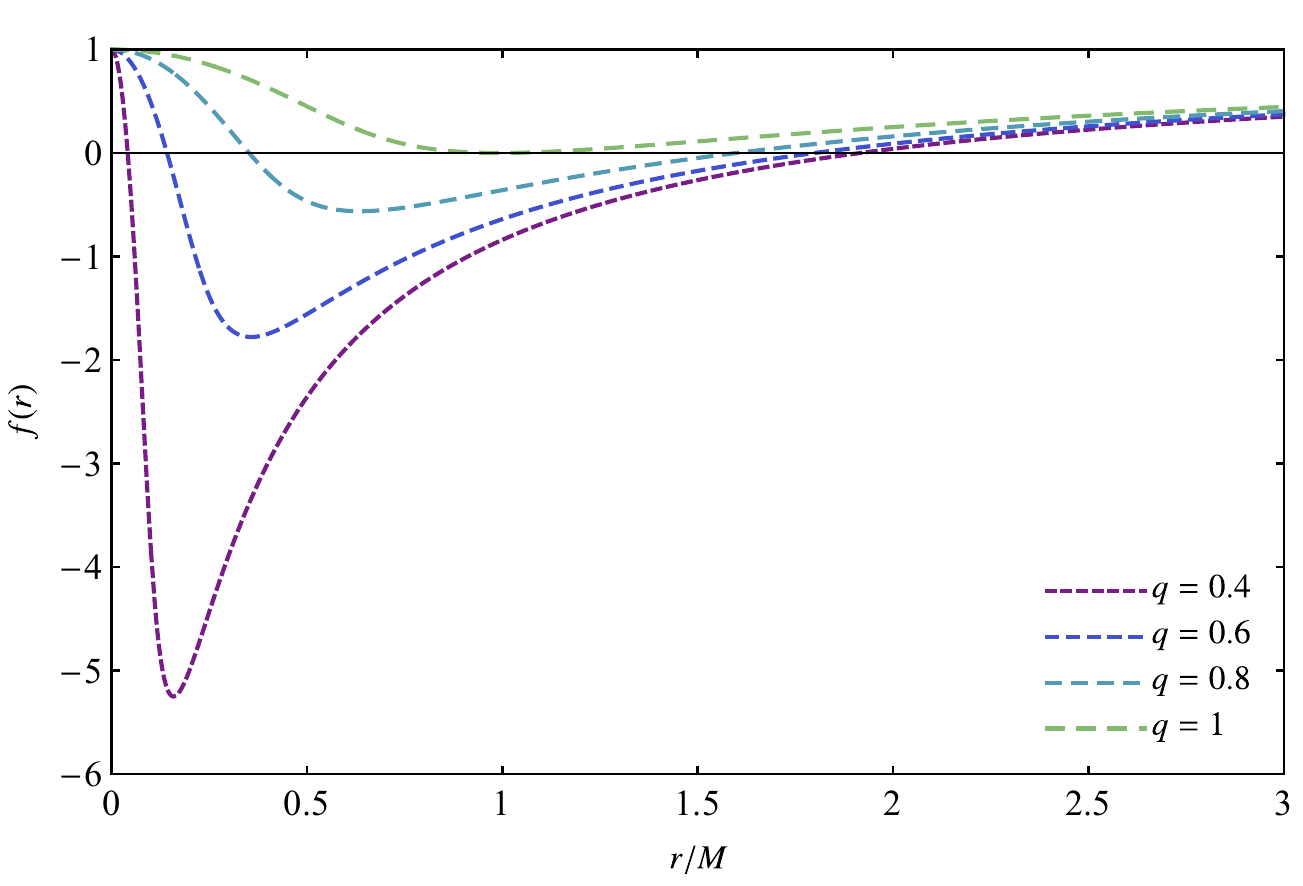}
		\caption{\footnotesize{Metric function~\eqref{n84a}, for distinct values of $q$, as a function of $r/M$.}}
		\label{mf0}
	\end{centering}
\end{figure}

The metric function given by Eq.~\eqref{n84a} has the following $x=0$ and $x\to\infty$ behaviors:
\begin{align}
	\label{n85}	f(r) & = 1- \dfrac{432 \left(3-2 \sqrt{2}\right) M^4}{\pi ^3 Q^6}r^{2}+\mathcal{O}\left[r^5\right], \\
	\label{n86}	f(r) & = 1-\dfrac{2M}{r}+\dfrac{Q^2}{r^{2}}-\dfrac{\left(17+12 \sqrt{2}\right) \pi ^8 Q^{18}}{181398528 M^8 r^{10}}+\mathcal{O}\left[\dfrac{1}{r^{11}}\right]\,,
\end{align}
respectively. In Fig.~\ref{mf0}, we display the behavior of the metric function~\eqref{n84a}. We see that for $0 < q < 1$, we have RBH solutions with an event horizon $r_{+}$ and a Cauchy horizon $r_{-}$. The location of the horizons and the extremal charge value can be found by solving $f(r) = 0$ and $f^{\prime}(r) = 0$ numerically. When $q = 1$, we obtain an extremally charged RBH geometry, with $r_{\text{ext}} \approx 0.9925M$ and $Q_{\text{ext}} \approx  1.0009M$ being the location of the event horizon and the extremal charge value, respectively. For $q > 1$, we have horizonless solutions, whose scope is beyond this paper, and the Schwarzschild BH solution is obtained when $q = 0$. This causal structure is similar to several nonsingular charged BH geometries presented in the literature (see, e.g., Ref.~\cite{Bronnikov:2022ofk} and references therein).

As discussed in Refs.~\cite{Lobo:2020ffi,Bronnikov:2012wsj}, for an SSS spacetime, the Kretschmann scalar (KS) is sufficient to ensure that the geometry is regular, in the sense that all invariants constructed from the Riemann tensor and the metric tensor are finite. The KS is defined by $\mathcal{K}_{\text{S}} \equiv R_{\mu\nu\rho\sigma}R^{\mu\nu\rho\sigma}$, where $R_{\mu\nu\rho\sigma}$ are the covariant components of the Riemann tensor. Using Eq.~\eqref{2}, we find that~\footnote{As far as we know, there are no regular metrics that are solutions of GR minimally coupled to LED, see~\ref{appx2} for details.}
\begin{equation}
	\label{KS}\mathcal{K}_{\text{S}}=\frac{4 \left(1-f(r)\right)^2+4 r^2 \left(f(r)^{\prime}\right)^2+r^4 \left(f(r)^{\prime\prime}\right)^2}{r^4}\,.
\end{equation}

\begin{figure}[!htbp]
	\begin{centering}
		\includegraphics[width=0.6\columnwidth]{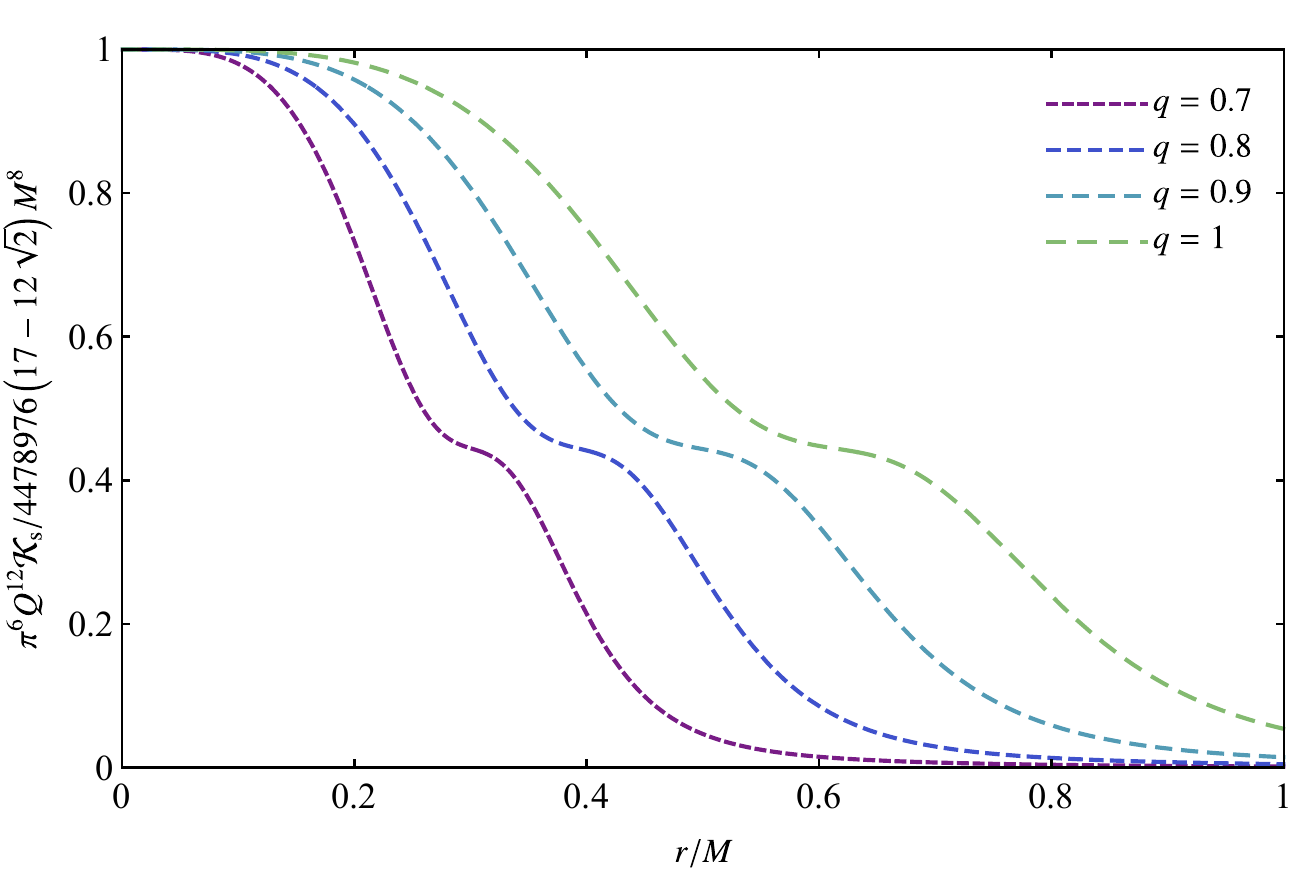}
		\caption{\footnotesize{KS of the spacetime given by the metric function~\eqref{n84a}, considering different choices of $q$, as a function of $r/M$.}}
		\label{figks0}
	\end{centering}
\end{figure}

The behavior of the corresponding KS is shown in Fig.~\ref{figks0}. We note that KS is finite in the limit $r \rightarrow 0$, namely,
\begin{equation}
	\mathcal{K}_{\text{S}}(r) = \dfrac{4478976 \left(17-12 \sqrt{2}\right) M^8}{\pi ^6 Q^{12}} + \mathcal{O}\left[r^{4}\right],
\end{equation}
which indicates that as long as $|Q| > 0$, the geometry is regular for $r \ge 0$.

%%%%%%%%%%%%%%%%%%%%%%%%%%%%%%%%%%%%%%%%%%%%%%%%%%
\subsection{Fully RBH spacetimes with h = 2}\label{sech2}
%%%%%%%%%%%%%%%%%%%%%%%%%%%%%%%%%%%%%%%%%%%%%%%%%%

For $h=2$, the asymptotically flat metric corresponds to the following NED Lagrangian
\begin{equation}
	\label{n832}\mathcal{L}(x)=\frac{2048 M^4}{\pi ^4 Q^6}\, \frac{1- x^4}{(1+x^4)^2}\,,
\end{equation}
where
\begin{equation}
	\label{h2K}K:=\frac{\pi Q^2}{4\sqrt{2}M}\,.	
\end{equation}
The metric function $f(x)$ takes the form~\cite{Azreg-Ainou:2025tuj}
\begin{equation}
	\label{n84a2}f(x)=1-\frac{32 M^2}{\pi^2 Q^2}~p(x)\,,	
\end{equation}
and
\begin{align}
	\label{n84b2}p(x) = & \ \frac{\sqrt{2}}{24 x} \Big[6  [\arctan (1+\sqrt{2} x)-\arctan (1-\sqrt{2} x)]+ 3\ln \left(\frac{1-\sqrt{2}x+x^2}{1+\sqrt{2} x+x^2}\right)\Big]\,.
\end{align}

\begin{figure}[!htbp]
	\begin{centering}
		\includegraphics[width=0.6\columnwidth]{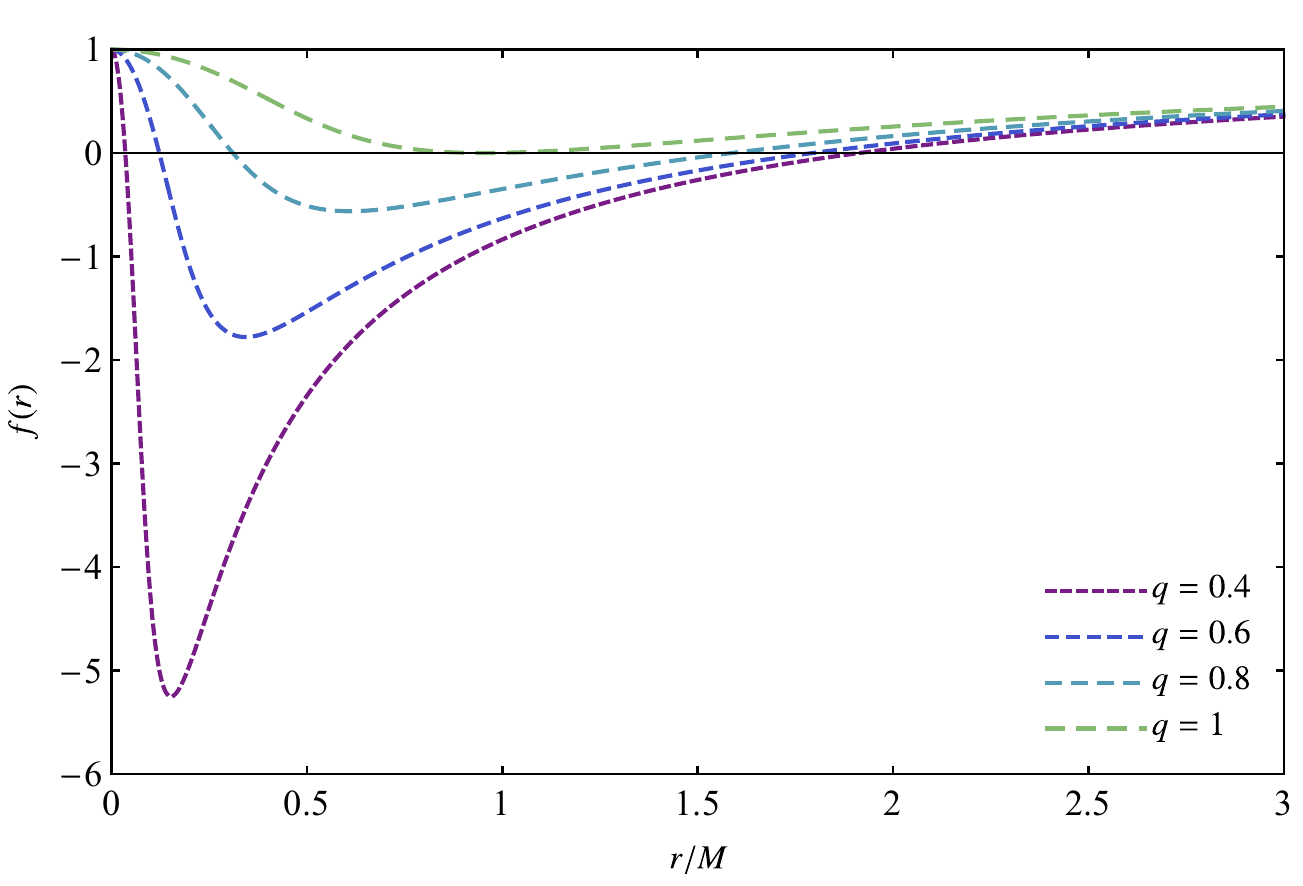}
		\caption{\footnotesize{Metric function~\eqref{n84a2}, for distinct values of $q$, as a function of $r/M$.}}
		\label{mf}
	\end{centering}
\end{figure}

At the asymptotic limits $x=0$ and $x\to\infty$, the metric function~\eqref{n84a2} has the following behavior
\begin{align}
	\label{n852}f(r) & = 1-\frac{1024 M^4}{3 \pi ^4 Q^6} r^2+\mathcal{O}\left[r^3\right], \\
	\label{n862}f(r) & = 1-\dfrac{2M}{r}+\dfrac{Q^2}{r^{2}}-\dfrac{\pi ^4 Q^{10}}{5120 M^4 r^6}+\mathcal{O}\left[\dfrac{1}{r^{7}}\right],
\end{align}
respectively. In Fig.~\ref{mf}, we show the behavior of the metric function~\eqref{n84a2}. The causal structure of this spacetime is similar to that of the case given by Eq.~\eqref{n84a}, i.e., BH solutions are described by $0 \leq q \leq 1$. However, the extremally charged BH is characterized by $r_{\text{ext}} \approx 0.9526M$ and $Q_{\text{ext}} \approx  1.0108M$. We also note that the class $h = 2$ more closely resembles the RN than the class $h = 0$. This can be seen from the asymptotic expansions of the corresponding metric functions [see Eqs.~\eqref{n86} and~\eqref{n862}, respectively]. The first higher-order corrections for the case $h = 0$ in the far field appear only at the order of $1/r^{10}$, whereas for $h = 2$, we can already observe corrections at the order of $1/r^{6}$. This difference plays an important role when seeking corrections to the NED in the weak-field limit. In this particular case, it indicates that it is easier to find NED corrections in the weak-field regime for the case $h = 2$ than for the case $h = 0$.

\begin{figure}[!htbp]
	\begin{centering}
		\includegraphics[width=0.6\columnwidth]{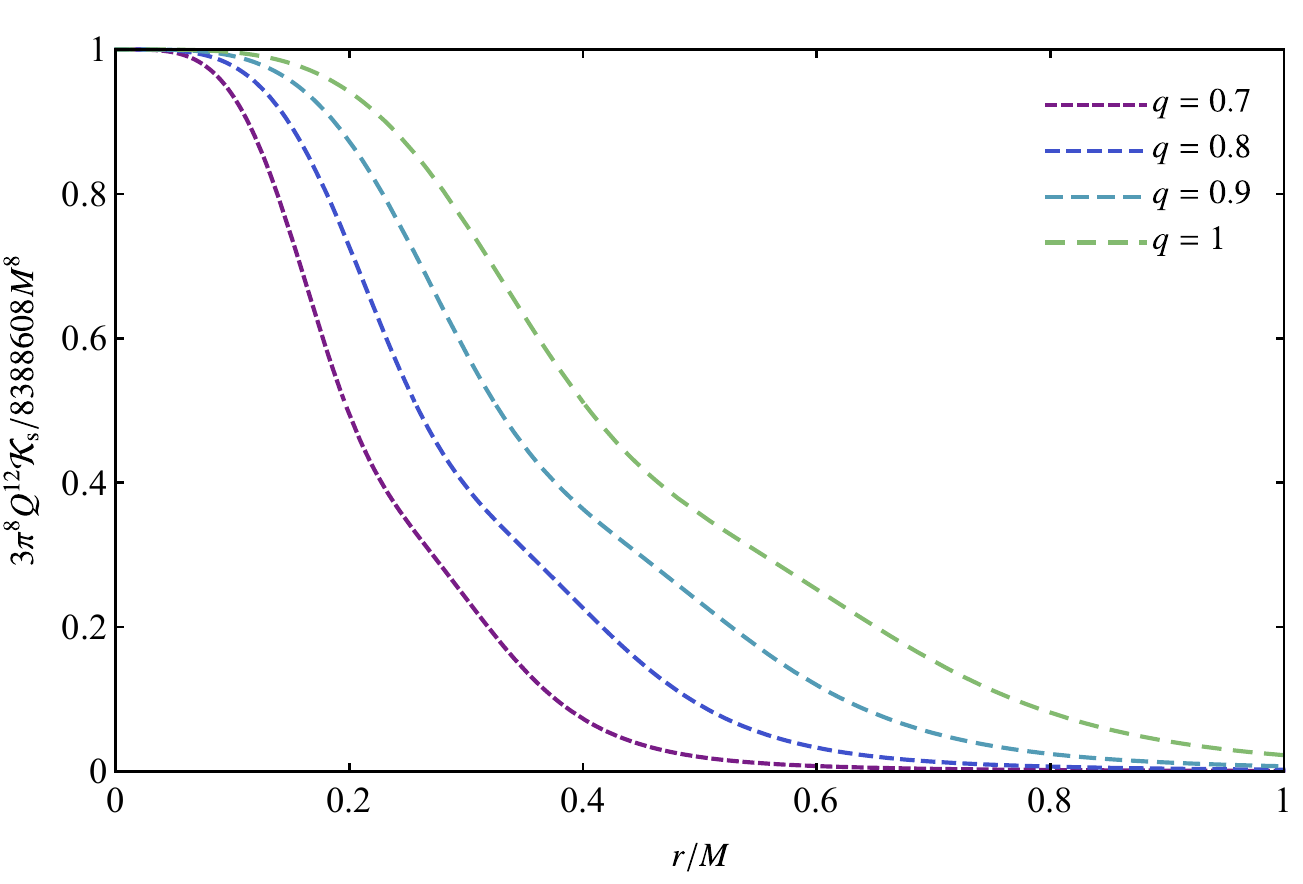}
		\caption{\footnotesize{KS of the spacetime given by the metric function~\eqref{n84a2}, considering different choices of $q$, as a function of $r/M$.}}
		\label{ks}
	\end{centering}
\end{figure}

The behavior of the corresponding KS is shown in Fig.~\ref{ks}. As we can see, the KS is finite in the limit $r \rightarrow 0$, i.e.,
\begin{equation}
	\mathcal{K_{\text{S}}}(r) = \dfrac{8388608M^{8}}{3 \pi ^8 Q^{12}}+ \mathcal{O}\left[r^{2}\right].
\end{equation}
Therefore, as long as $|Q| > 0$, the geometry is regular.

We also point out that the de Sitter-like behavior of the metric functions~\eqref{n84a} and~\eqref{n84a2} at their core (see Eqs.~\eqref{n85} and~\eqref{n852}, respectively) is an expected feature of NED-sourced spacetimes that satisfy the weak energy condition~\cite{Dymnikova:2004zc}. In particular, we can obtain the “effective cosmological constant” from the asymptotic analysis of the metric functions at the core [see, for example, equations ~\eqref{n85} and ~\eqref{n852}]. This can be done by comparing the asymptotic form of these metric functions with the metric function of the de Sitter spacetime, namely,
\begin{equation}
	\label{deSitter}f(r) = 1 - \dfrac{\Lambda}{3}r^{2},
\end{equation}
where $\Lambda$ is the cosmological constant. Our results indicate that the effective cosmological constant for the $h = 2$  case is nearly $46\%$ larger than that for the $h = 0$ case, regardless of the value of $Q/M$. This may have some interesting consequences for analyses of the inner horizon structure, as well as of the thermodynamic properties, but it is beyond the scope of this work.

Moreover, since the spacetimes of the h-family discussed here satisfy a correspondence with Maxwell's theory in the far-field, i.e.,
\begin{equation}
	\label{maxwelasym}\mathcal{L}(F) \rightarrow F \quad \text{and} \quad \mathcal{L}_{F} \rightarrow 1,
\end{equation}
for $F \ll 1$, their metric functions behave as the RN metric function for large $r$, which is given by
\begin{equation}
	f^{\text{RN}}(r) \equiv 1-\dfrac{2M}{r}+\dfrac{Q^2}{r^{2}},
\end{equation}
with higher-order corrections. For the RN spacetime, the 
extremally charged BH is characterized by $r_{\text{ext}} = Q_{\text{ext}} = M$. It is also interesting to note that the KS function profile for both geometries shows a notable difference. For the case $h=0$, we observe the formation of a “nearly flat region” in the KS function profile (see Fig.~\ref{figks0}), whereas this does not appear to be the case for $h = 2$, as shown in Fig.~\ref{ks}. This indicates that, in these regions, the derivative of the KS function with respect to the radial coordinate $r$ is close to zero. We can understand this by noting that the functions $p(x)$ for $h = 0$ and $h = 2$ are very different [see equations $\eqref{n84b}$ and $\eqref{n84b2}$, respectively]. The $h=0$ case contains trigonometric sums and arctangents that create a “step-like” transition in the derivatives of the metric function, while the simpler logarithmic and arctangent terms for the $h=2$ case result in a smoother, monotonic decrease in curvature. These distinct behaviors may be relevant to the study of physical effects such as tidal forces, which strongly depend on the Riemann tensor. 

\begin{figure}[!htbp]
	\begin{centering}
		\includegraphics[width=0.6\columnwidth]{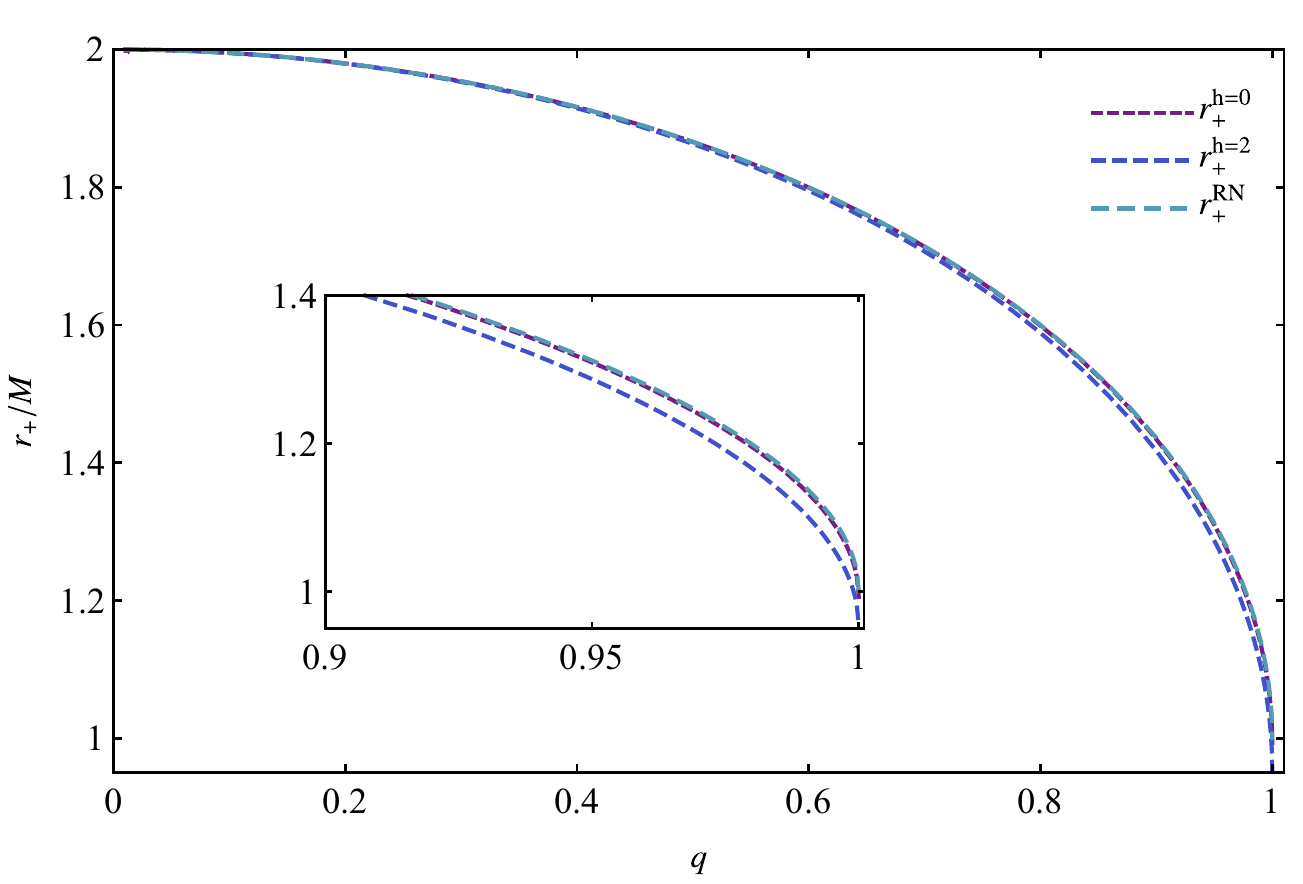}
		\caption{\footnotesize{The event horizon location for fully RBHs and RN BHs, as a function of $q$.}}
		\label{horizons}
	\end{centering}
\end{figure}

For completeness, in Fig.~\ref{horizons}, we discuss the locations of the event horizons for the fully RBHs compared to that of the RN case. As we can see, the location of the horizons for both fully RBHs and RN BH, with a fixed $q$, are very similar. The similarity of the horizons implies that the area of these BH solutions will also be quite similar, since the area is given by $A = 4 \pi r^{2}_+$. However, the similarity of the areas does not imply that the phenomenology associated with the solutions is the same. A notable example is the Ayón-Beato-García (ABG) RBH solution~\cite{Ayon-Beato:1998hmi}. The ABG and RN BH solutions also have very similar event horizon locations for a fixed value of the normalized electric charge, but the ABG solution gives rise to several phenomena that do not occur in the RN solution, such as unbounded absorption~\cite{dePaula:2024xnd} and superradiant instability~\cite{Dolan:2024qqr}.

\section{Motion of photons}\label{sec:mp}

In this section, we investigate the motion of photons in NED. Due to the spherical symmetry of the spacetime~\eqref{sss0}, we consider the motion in the equatorial plane, i.e., $\theta = \pi/2$, without loss of generality.

%%%%%%%%%%%%%%%%%%%%%%%%%%%%%%%%%%%%%%%%%%%%%%%%%%
\subsection{Circular null geodesics of the effective geometry (EG)}\label{sech2mp}
%%%%%%%%%%%%%%%%%%%%%%%%%%%%%%%%%%%%%%%%%%%%%%%%%%

As discussed in the Introduction, the motion of photons in NED is not described by the standard geometry (SG), but rather by the EG~\cite{Novello:1999pg}.
The EG metric can be written as~\cite{Bronnikov:2000vy}
\begin{align}
	\label{eg1}\bar{g}^{\mu\nu}=\mathcal{L}_F g^{\mu\nu} - 4 \mathcal{L}_{FF}F^\mu{}_\alpha F^{\alpha\nu}\,,
\end{align}
and the corresponding line element results in
\begin{align}
	\dd \bar{s}^2 & \equiv \bar{g}_{\mu\nu}\dd x^{\mu}\dd x^{\nu} = \frac{1}{\Phi}\left(f(r)\dd t^2 - \frac{1}{f(r)}\dd r^2\right) - \frac{r^2}{\mathcal{L}_F}\,\dd\Omega^2\,,
\end{align}
where $\Phi = \mathcal{L}_F + 2F\mathcal{L}_{FF}$. For LED, we have $\bar{g}^{\mu\nu}=g^{\mu\nu}$.

Following the same approach presented in Refs.~\cite{dePaula:2023ozi,dePaula:2024yzy}, we apply the Hamiltonian formalism in curved spacetimes to analyze the general geodesic motion in the EG, then specialize to circular geodesics in the EG. The Hamiltonian that provides the equations of motion for photons in the EG is given by
\begin{align}
	\label{H_EG} H & \equiv \dfrac{1}{2} \bar{g}^{\mu\nu}\bar{p}_{\mu}\bar{p}_{\nu} = \dfrac{1}{2}\left[\Phi\left(\dfrac{\bar{p}_{t}^{2}}{f(r)} - f(r)\,\bar{p}_{r}^{2}\right) - \dfrac{\mathcal{L}_{F}\bar{p}_{\varphi}^{2}}{r^{2}}\right],
\end{align}
where $\bar{p}_\mu$ ($p_{\mu}$) are the components of the four-momentum of photons (massless particles). Since $\bar{p}_{\mu} = \bar{g}_{\mu\nu}\dot{x}^{\nu}$, we find that
\begin{align}
	\label{eqm1_EG}\dot{t} & = \dfrac{\Phi E_{\text{ph}}}{f(r)}  ,\\
	\label{eqm2_EG}\dot{r} & = - f(r)\Phi \bar{p}_{r}  ,\\
	\label{eqm3_EG}\dot{\varphi} & =  \dfrac{\mathcal{L}_{F} L_{\text{ph}}}{r^{2}},
\end{align}
for which the overdot stands for differentiation with respect to an affine parameter. Due to the symmetries of the Hamiltonian~\eqref{H_EG} with respect to the coordinates $t$ and $\varphi$, we take $\bar{p}_{t} \equiv E_{\text{ph}}$ and $\bar{p}_{\varphi} \equiv -L_{\text{ph}}$ as constants of motion, where $E_{\text{ph}}$ and $L_{\text{ph}}$ are the energy and angular momentum of photons, respectively. To avoid misinterpretation, we have added the subscript ``ph'' to the energy and angular momentum of the photon.

Using Eqs.~\eqref{H_EG}-\eqref{eqm3_EG} and $H = 0$, we find that
\begin{equation}
	\label{RE_EG}\Big(\dfrac{\dot{r}}{\Phi}\Big)^{2} + V(r) = E_{\text{ph}}^{2},
\end{equation}
where $V(r)$ is defined as
\begin{equation}
	\label{Veff_EG}V(r) \equiv L_{\text{ph}}^{2}\dfrac{\mathcal{L}_{F} f(r)}{\Phi r^{2}}.
\end{equation}

Closed circular photon orbits in the EG are described by~\eqref{RE_EG} with $\dot{r} = 0$ and $\ddot{r} = 0$, which imply that $V(r)=E_{\text{ph}}^{2}$  and $V'(r) = 0$, respectively. Moreover, if $V''(r)<0$, then the closed circular photon orbit is unstable. From $\dot{r} = 0$, we obtain the critical impact parameter associated with the light ring (LR), namely
\begin{equation}
	\label{CIP_EG}b_{l} =  r_{l}\sqrt{\dfrac{\Phi_{l}}{\left(\mathcal{L}_{F}\right)_{l} f_{l}}},
\end{equation}
where $f(r_{l}) \equiv f_{l}$ and $r_l$ is the radial coordinate of the closed circular photon orbit. The parameter $b_l$ defines a transition boundary separating absorbed null geodesics ($b < b_l$) from scattered null geodesics ($b > b_l$). Moreover, from $\ddot{r} = 0$, we obtain the corresponding radial coordinate of the LR, given by
\begin{equation}
	\label{CR_EG}f_{l}\left[2-\dfrac{r_{l}\left(\left(\mathcal{L}_{F}\right)_{l}\right)^{\prime}}{\left(\mathcal{L}_{F}\right)_{l}} + \dfrac{r_{l}\left(\Phi_{l}\right)^{\prime}}{\Phi_{l}} \right]-r_{l}f_{l}^{\prime} = 0,
\end{equation}
where the subscript ``$l$'' denotes that the quantity under consideration is computed at the radial coordinate of the LR $r_{l}$. If the NED model satisfies a correspondence with Maxwell's theory [see the condition given by Eq.~\eqref{maxwelasym}], Eqs.~\eqref{RE_EG}-\eqref{CR_EG} reduce to the corresponding results for massless particles~\cite{Paula:2020yfr}. Thus, we find that
\begin{equation}
	\label{RE_SG}\dot{r}^{2} + U(r) = E^{2},
\end{equation}
where $E$ is the energy of the massless particle, and $U(r)$ is given by
\begin{equation}
	\label{Veff_SG}U(r) \equiv L^{2}\dfrac{f(r)}{r^{2}},
\end{equation}
with $L$ being the angular momentum of the massless particle, and
\begin{equation}
	\label{bcrc}b_{c} =  \dfrac{r_{c}}{\sqrt{f_{c}}} \quad \text{and} \quad 2f_{c}-r_{c}f_{c}^{\prime} = 0.
\end{equation}

In this case, $\mathcal{L}_{F} \rightarrow 1$ and $\Phi \rightarrow 1$ [cf. Eq.~\eqref{maxwelasym} and recall that $\Phi = \mathcal{L}_F + 2F\mathcal{L}_{FF}$], and instead of the LR coordinate, we have a critical radius $r_{c}$ associated with the critical impact parameter $b_{c}$. In LED, these results describe the motion of photons, but in NED, they describe the motion of massless particles of a nature other than electromagnetic. Examples of such particles include gravitons, from a hypothetical perspective, but we can also effectively consider high-energy neutrinos as massless particles (see, e.g., Ref.~\cite{Stuchlik:2019uvf} and references therein).

We note that concerning the perspective of static observers in NED-sourced spacetimes, the NED fields introduce non-trivial contributions to the motion of photons. Consequently, for an observer placed at an arbitrary location outside the event horizon, the critical impact parameter will not necessarily correspond to the shadow radius~\cite{dePaula:2023ozi}. To properly analyze this problem, we can investigate the orbit equation~\eqref{RE_EG}.

Using Eqs.~\eqref{eqm3_EG} and~\eqref{RE_EG}, we can show that
\begin{equation}
	\label{OQ2}\dfrac{\dd r}{\dd \varphi} = \pm r\sqrt{\dfrac{f(r)\Phi}{\mathcal{L}_{F}}\Big[\dfrac{r^{2}\Phi E_{\text{ph}}^{2}}{f(r)\mathcal{L}_{F}L_{\text{ph}}^{2}} - 1\Big]}.
\end{equation}
Moreover, the turning point, $R$, of the photon orbit, i.e., the radius of the nearest point, is given by
\begin{equation}
	\label{tp}\dfrac{\dd r}{\dd \varphi}\bigg|_{r=R} = 0.
\end{equation}
Combining Eqs.~\eqref{OQ2} and~\eqref{tp}, we can find an expression for $E_{\text{ph}}/L_{\text{ph}}$ at the turning point $R$ and rewrite Eq.~\eqref{OQ2} as
\begin{equation}
	\label{OQ3}\dfrac{\dd r}{\dd \varphi} = \pm r\sqrt{\dfrac{f(r)\Phi}{\mathcal{L}_{F}}\Big[\dfrac{r^{2}\Phi f_{R}\mathcal{L}_{F}(R)}{R^{2}\Phi_{R}f(r)\mathcal{L}_{F}} - 1\Big]}.
\end{equation}

If we consider a static observer placed at $r_{\text{o}}$ sending a light ray that is transmitted into the past with an angle $\beta$ with respect to the radial direction, we obtain
\begin{align}
	\label{tst} \cot\beta = \dfrac{\bar{g}_{rr}}{\bar{g}_{\varphi\varphi}}\sqrt{\dfrac{g_{\varphi\varphi}}{g_{rr}}}\dfrac{\dd r}{\dd \varphi}\bigg|_{r = r_{\text{o}}}  = \dfrac{\mathcal{L}_{F}}{r f(r)^{1/2}\Phi}\dfrac{\dd r}{\dd \varphi}\bigg|_{r = r_{\text{o}}}\,.
\end{align}
The corresponding expression of Eq.~\eqref{tst} in LED was derived in~\cite{Belhaj:2020nqy}. Injecting Eq.~\eqref{OQ3} into Eq.~\eqref{tst}, and using $\sin^{2}\beta = 1/(1+\cot^{2}\beta)$, we find that the shadow radius of electrically charged NED-sourced BHs is given by
\begin{equation}
	\label{sr} \bar{r}_{s} = r_{\text{o}}\sin\beta = r_{\text{o}}b_{l}\sqrt{\dfrac{f_{\text{o}} \Phi_{\text{o}}}{r_{\text{o}}^{2}\Phi_{\text{o}}+b_{l}^{2}f_{\text{o}}\left(\Phi_{\text{o}}-\mathcal{L}_{F}(r_{\text{o}}) \right)}},
\end{equation}
where $b_l$ is given in~\eqref{CIP_EG} in terms of $r_{l}$, which is the radius of the photon sphere solution to Eq.~\eqref{CR_EG}. We point out that $\Phi_{\text{o}}-\mathcal{L}_{F}(r_{\text{o}})$ in Eq.~\eqref{sr} can be negative for $h = 0$ if
\begin{equation}
	r < r_{\text{eff}}^{h = 0} \equiv \dfrac{\sqrt{1+\sqrt{2}} \pi  Q^2}{2 \sqrt[4]{2} 3^{7/8} M},
\end{equation}
where $r_{\text{eff}}$ is an effective radius, and for the case $h = 2$, if
\begin{equation}
	r < r_{\text{eff}}^{h = 2} \equiv \dfrac{\sqrt[4]{3} \pi  Q^2}{4 \sqrt{2} M}.
\end{equation}
However, one can show that $r_{\text{eff}}^{h = 0}$ and $r_{\text{eff}}^{h = 2}$ are always located within the BH, i.e., $r_{+} > r_{\text{eff}}$. Therefore, we conclude that the shadow radius is not affected by this behavior. We also point out that the sign flip of a spacetime metric can be effectively explored as a tool to prevent the occurrence of curvature singularities in certain physical setups~\cite{Capozziello:2024ucm, Capozziello:2025wwl}.

Notice also that if we place the static observer very far away from the BH, i.e., for $r_{\text{o}} \rightarrow \infty$, we have
\begin{equation}
	\label{sr_NED}\bar{r}_{s} = b_{l},
\end{equation}
as expected for static and spherically symmetric spacetimes. We emphasize that the result presented by Eq.~\eqref{sr_NED} works for fully RBHs with $h = 0$ and $h = 2$, but may not be valid for arbitrary NED sourced spacetimes. Moreover, we denote the ``shadow radius'' computed according to the SG as $r_{s}$.

\begin{figure}[!htbp]
	\begin{centering}
		\includegraphics[width=0.5\columnwidth]{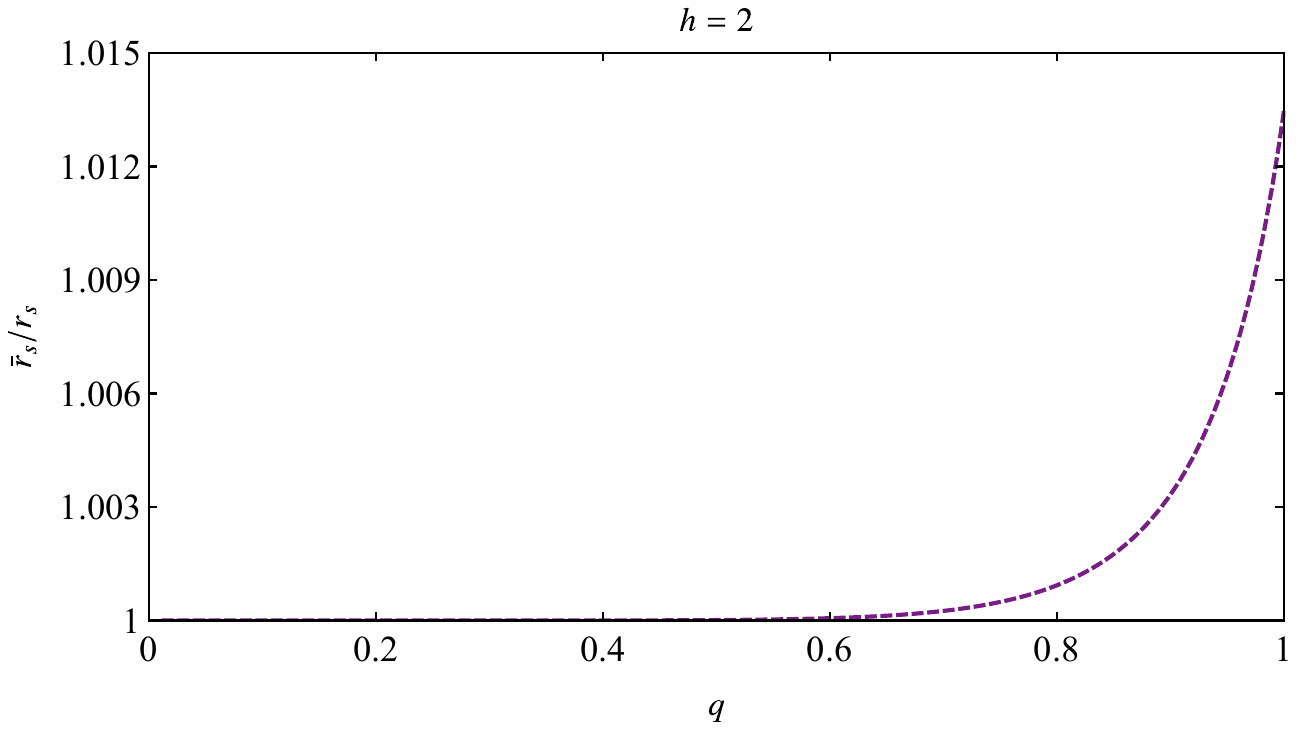}
		\includegraphics[width=0.5\columnwidth]{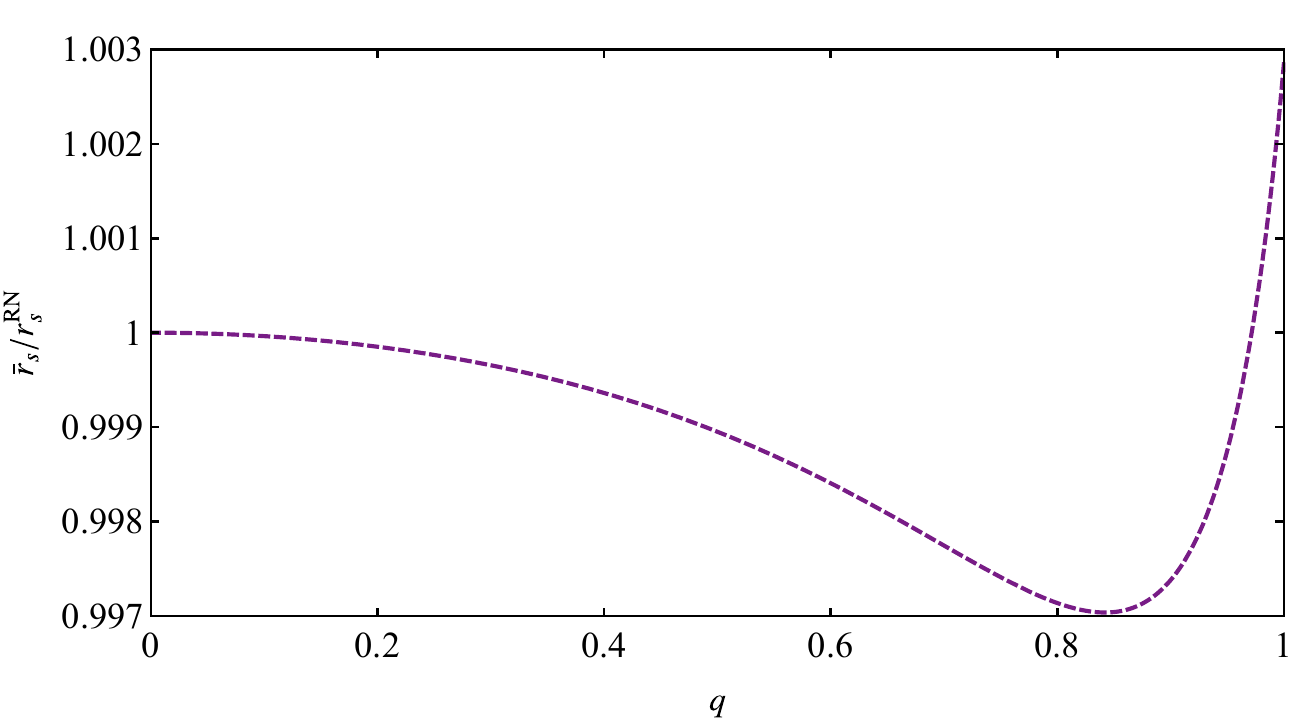}
		\caption{\footnotesize{Ratio between the shadow radius of the fully RBH with $h = 2$, as a function of $q$, for two distinct scenarios: (i) considering the effective $\bar{r}_{\text{s}}$ and standard $r_{\text{s}}$ geometries (top panel); and (ii) considering $\bar{r}_{\text{s}}$ and the shadow radius of the RN case, denoted by $r_{\text{s}}^{\text{RN}}$ (bottom panel).}}
		\label{figsr}
	\end{centering}
\end{figure}

The ratio between the shadow radius, considering the effective and standard geometries, can help us quantify how much the EG affects the shadow shape compared to the SG. For $h=2$, as shown in Fig.~\ref{figsr}, we see that $\bar{r}_{\text{s}}$ is always greater than $r_{\text{s}}$, provided, of course, that $Q > 0$. The largest difference between them occurs for the extremal BH case, where $\bar{r}_{\text{s}} \approx 1.01346r_{\text{s}}$. Moreover, the comparison with the RN case shows that the shadow radius of the RBH with $h = 2$ is smaller (greater) than that of the RN case for $q < q_{\text{crit}} \cong 0.971611$ ($q > q_{\text{crit}}$).

\begin{figure}[!htbp]
	\begin{centering}
		\includegraphics[width=0.6\columnwidth]{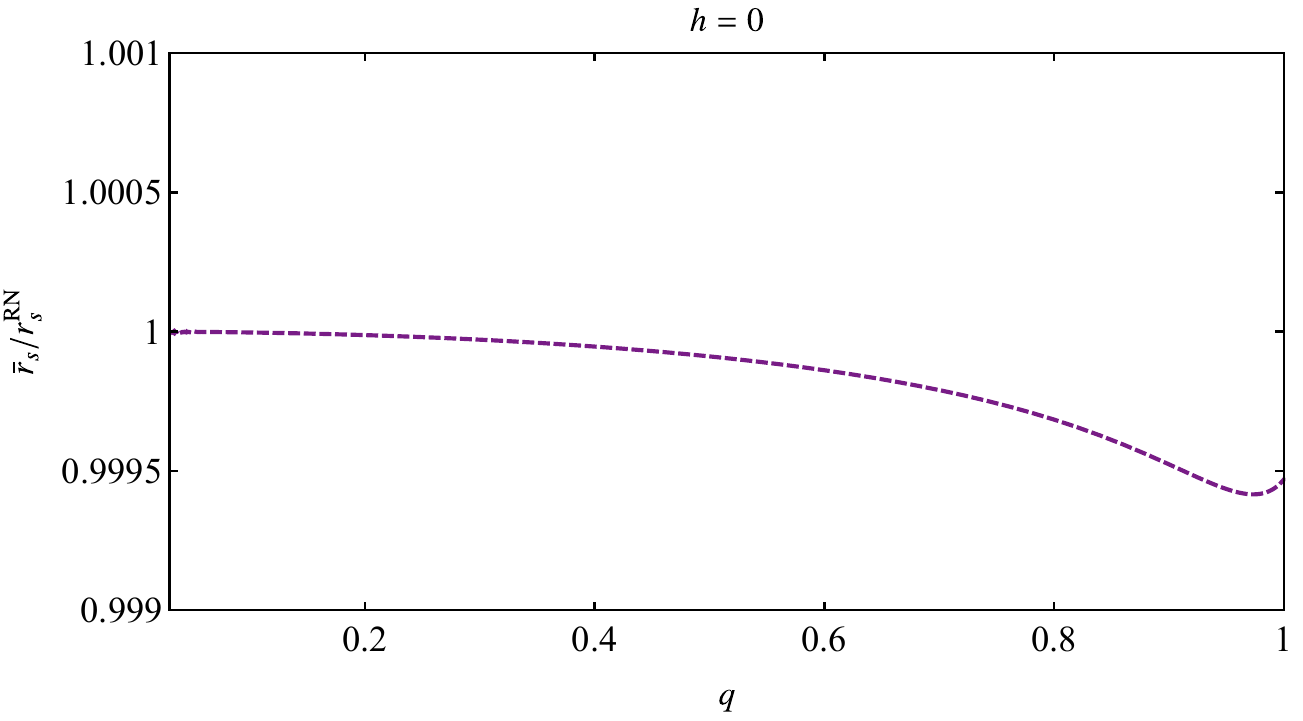}
		\caption{\footnotesize{Ratio between the shadow radius of the fully RBH with $h = 0$, as a function of $q$, considering $\bar{r}_{\text{s}}$ and the shadow radius of the RN case $r_{\text{s}}^{\text{RN}}$.}} 
		\label{figsr2}
	\end{centering}
\end{figure}

The results for $h = 0$ on the ratio between the shadow radius computed considering the standard and effective geometries are qualitatively similar to those of the case $h = 2$. The highest difference occurs for the extremal BH case, for which $\bar{r}_{\text{s}} \approx 1.00284r_{\text{s}}$. Nevertheless, compared to the RN case, the results for $h = 0$ are different from those obtained for the case $h = 2$. As shown in Fig.~\ref{figsr2}, for the case $h = 0$, the corresponding shadow radius is typically smaller than that for the RN case regardless of the value of $q$.

Therefore, for fully RBH solutions with $h = 0$ or $h = 2$~\cite{Azreg-Ainou:2025tuj}, the difference between the shadow radius considering the standard and effective geometries is very small. This result is in contrast to those obtained for other NED-sourced BH geometries~\cite{dePaula:2023ozi}. Despite the similarity, the EG is still necessary to calculate the proper motion of photons~\cite{dePaula:2024yzy,Russo:2024llm,Murk:2024nod}.

%%%%%%%%%%%%%%%%%%%%%%%%%%%%%%%%%%%%%%%%%%%%%%%%%%
\subsection{Weak deflection angle}\label{sech2wd}
%%%%%%%%%%%%%%%%%%%%%%%%%%%%%%%%%%%%%%%%%%%%%%%%%%

Here, we use the Gauss-Bonnet theorem~\cite{Gibbons:2008rj} (see, e.g., Refs.~\cite{Ovgun:2019wej,Javed:2019kon,Javed:2020lsg,Fu:2021akc,Okyay:2021nnh}, and references therein for more details) to derive an expression for the weak deflection angle of the $h = 0$ and $h = 2$ family of fully RBH solutions in the weak field limit, considering the standard and effective geometries.

The line element of the optical metric related to the motion of massless particles can be obtained by setting $\dd s^{2} = 0$ and $\theta = \pi/2$ in the line element~\eqref{2}. Thus, we get
\begin{equation}
	\label{LE_OM2}\dd t^{2} = h_{\mu\nu}\dd x^{\mu}\dd x^{\nu} = \dd r_{\star}^{2}+v(r_{\star})^{2}\dd \varphi^{2},
\end{equation}
where $r_{\star}$ is the Regge–Wheeler coordinate, defined by $\dd r_{\star} = f(r)^{-1}\dd r$, and $v(r_{\star}) \equiv r/\sqrt{f(r)}$. The corresponding Ricci scalar $R$ associated with this geometry is given by
\begin{equation}
	R = -\dfrac{2}{v(r_{\star})}\dfrac{d^{2}v(r_{\star})}{dr_{\star}^{2}},
\end{equation}
and the Gaussian optical curvature $\mathcal{K}$ yields $\mathcal{K} = R/2$.

The Gauss-Bonnet theorem allows us to relate the deflection angle $\Theta(b)$ to the Gaussian optical curvature $\mathcal{K}$ through a simple equation, given by~\cite{Gibbons:2008rj}
\begin{equation}
	\label{DA_GBT}\Theta(b) = - \int_{0}^{\pi + \alpha}\int_{u(\varphi)}^{\infty}\mathcal{K}\dd S,
\end{equation}
where
\begin{equation}
	\dd S = \sqrt{|h|}\dd r_{\star}\dd \varphi,
\end{equation}
with $h$ being the determinant of the metric tensor $h_{\mu\nu}$. The parameters $\alpha$ and $u(\varphi)$ are corrections to the observer's position and the trajectory of the particle, respectively. They can be obtained by solving the radial orbit equation. For the general SSS metric given by Eq.~\eqref{sss0}, we have 
\begin{equation}\label{diff}
	\left(\dfrac{\dd u}{\dd \varphi}\right)^{2}+ u^{4} C(r)^{2}\left(\dfrac{1}{A(r)B(r)b^{2}}+\dfrac{1}{C(r)B(r)}\right) = 0,
\end{equation}
where we defined $r \equiv 1/u$, with $u = u(\varphi)$. Notice that by applying the same approach to the effective metric~\eqref{eg1}, the corresponding function $v(r_{\star})$ leads to
\begin{equation}
	\label{funcv}v^2_{\text{EG}}(r_{\star}) = \dfrac{r^{2}\Phi}{\mathcal{L}_{F} f(r)}.
\end{equation}
Hence, the equations we need to use to compute the deflection angle in the weak-field limit, considering the effective metric, are the same, provided that we replace $v(r_{\star})$ with $v_{\text{EG}}(r_{\star})$, and we take the metric components $A(r)$, $B(r)$, and $C(r)$ properly.

Considering the standard geometry, for our purposes, we can set the correction as $\alpha = 4M/b$, and suppose that $u(\varphi)$ is given by
\begin{equation}
	\label{pt}u(\varphi) = \dfrac{1}{b}\left[u_{0}(\varphi)+\dfrac{M}{b}u_{1}(\varphi) + \dfrac{M^{2}}{b^{2}}u_{2}(\varphi)\right],
\end{equation}
where the coefficients $u_{k}(\varphi)$, with $k = 0$, $1$, or $2$, can be obtained using an iterative method~\cite{Arakida:2011ty}. For fully RBH solutions with $h = 0$ and $h = 2$, the corresponding coefficients result in the same expressions since both exhibit Maxwell-like behavior in the far field. By inserting Eq.~\eqref{pt} into Eq.~\eqref{diff}, we obtain the following differential equation for $u_{0}(\varphi)$:
\begin{equation}
	\left(\dfrac{\dd u_{0}}{\dd \varphi}\right)^{2}+u_{0}^{2}-1 = 0,
\end{equation}
for which the physical solution is given by
\begin{equation}
	\label{u0}u_{0}(\varphi) = \sin\varphi.
\end{equation}
For $u_{1}(\varphi)$, we find that
\begin{equation}
	\cot\varphi \left(\dfrac{\dd u_{1}}{\dd \varphi}\right)+ u_{1}-\sin^{2}\varphi= 0,
\end{equation}
and the corresponding solution yields
\begin{equation}
	\label{u1}u_{1}(\varphi) = \left(1-\cos\varphi\right)^2.
\end{equation}
Finally, for $u_{2}(\varphi)$, we get
\begin{multline}
	M^2 \Big[\cos \varphi \dfrac{\dd u_{2}}{\dd \varphi} + \sin\varphi u_{2}\Big]+ \sin ^4\Big(\frac{\varphi }{2}\Big) \big[4 (Q^2-M^2) \cos \varphi \\+(3 M^2+Q^2) \cos 2 \varphi +M^2+3Q^2\big]= 0,
\end{multline}
and this yields
\begin{equation}
	\label{u2}u_{2}(\varphi) = \frac{6 \varphi  \left(Q^2-5 M^2\right) \cos \varphi +\sin \varphi  \left(-\left(3 M^2+Q^2\right) \cos (2 \varphi )+32 M^2 \cos
		\varphi +M^2-5 Q^2\right)}{8 M^2}.
\end{equation}
The corresponding expression for $u(\varphi)$ is obtained by inserting Eqs.~\eqref{u0},~\eqref{u1}, and~\eqref{u2} into Eq.~\eqref{pt}. We note that the equations for $u(\varphi)$, as well as the expansion $\mathcal{K} dS$ up to the fourth order in $r$, are the same for the cases $h = 0$ and $h = 2$. Moreover, at least up to $k = 2$, the results for the standard and effective geometries coincide. Therefore, one can show that the weak deflection angle of the standard and effective metrics, considering Eq.~\eqref{n84a} or Eq.~\eqref{n84a2}, coincides up to the third order in $1/b^{3}$, leading to
\begin{align}
	\Theta (b) =  \dfrac{4M}{b} + \dfrac{3\pi \left(5M^{2}-Q^{2}\right)}{4b^{2}} +\dfrac{16M\left(8M^{2}-3Q^{2}\right)}{3b^{3}} + \mathcal{O}\Big(\dfrac{1}{b^{4}}\Big),
\end{align}
which coincides with the RN result~\cite{Keeton:2005jd}, and for $Q = 0$ we obtain the Schwarzschild result. The similarity to the RN case can be understood by noting that the deviations from the RN metric first appear at orders $1/r^{10}$ and $1/r^{6}$, respectively, see the metric function given by Eq.~\eqref{n86}, and to $1/r^{6}$, see the metric function given by Eq.~\eqref{n862}. Consequently, the charge contributions are very small at the weak-field limit. To search for perturbations arising from the fully RBHs in the weak deflection angle, considering the standard and effective geometries, we need to consider more correction orders, which is beyond the scope of this work.

%%%%%%%%%%%%%%%%%%%%%%%%%%%%%%%%%%%%%%%%%%%%%%%%%%
\subsection{Astrophysical constraints\label{sech2ac}}
%%%%%%%%%%%%%%%%%%%%%%%%%%%%%%%%%%%%%%%%%%%%%%%%%%

In this section, we aim to impose astrophysical constraints on the BH charge-to-mass ratio $Q/M$ of the fully RBH solutions with $h = 0$ and $h = 2$ using observational data for Sgr A$^{\star}$ and M87$^{\star}$. The bounds in the shadow radius of Sgr A$^{\star}$~\cite{Vagnozzi:2022moj} and M87$^{\star}$~\cite{Chakhchi:2024tzo} consistent with the most updated observations at the $1\sigma$ and $2\sigma$ levels are summarized in Tab.~\ref{tab:bounds}. For Sgr A$^{\star}$, we consider the data from the instruments/teams Keck~\cite{Do:2019txf} and the Very Large Telescope Interferometer (VLTI)~\cite{GRAVITY:2020gka}, while for M87$^{\star}$, we consider the data from the Event Horizon Telescope (EHT) collaboration~\cite{EventHorizonTelescope:2019dse}. We also point out that the uncertainties in the mass-to-distance ratio of M87$^{\star}$ and Sgr A$^{\star}$ are incorporated into a parameter $\delta$, which measures the deviation between the infrared shadow radius and the Schwarzschild radius. Based on the value of $\delta$ and its uncertainties, the limits on the shadow radius of both galactic center BHs are determined. For further details, see, e.g., Eqs. (5)–(7) in Ref.~\cite{Vagnozzi:2022moj} for Sgr A$^{\star}$ and Tab. 3 in Ref.~\cite{Chakhchi:2024tzo} for M87$^{\star}$.
\begin{table}[hbtp!]
	\centering
	\begin{tabular}{||c|c|c||} 
		\hline
		\multirow{2}{2em}{BHs} & \multicolumn{2}{|c||}{Constraints} \\ \cline{2-3}
		& $1\sigma$ & $2\sigma$   \\ \hline
		Sgr A$^{\star}$ & $4.55 \lesssim r_{\text{s}}/M \lesssim 5.22$ & $4.21 \lesssim r_{\text{s}}/M \lesssim 5.56$   \\  \hline
		M87$^{\star}$ & $4.26 \lesssim r_{\text{s}}/M \lesssim 6.03$ & $3.38 \lesssim r_{\text{s}}/M \lesssim 6.91$ \\  \hline
	\end{tabular}
	
	\caption{Bounds in the shadow radius of Sgr A$^{\star}$ and M87$^{\star}$.}
	\label{tab:bounds}
\end{table} 

\begin{figure}[!htbp]
	\begin{centering}
		\includegraphics[width=0.6\columnwidth]{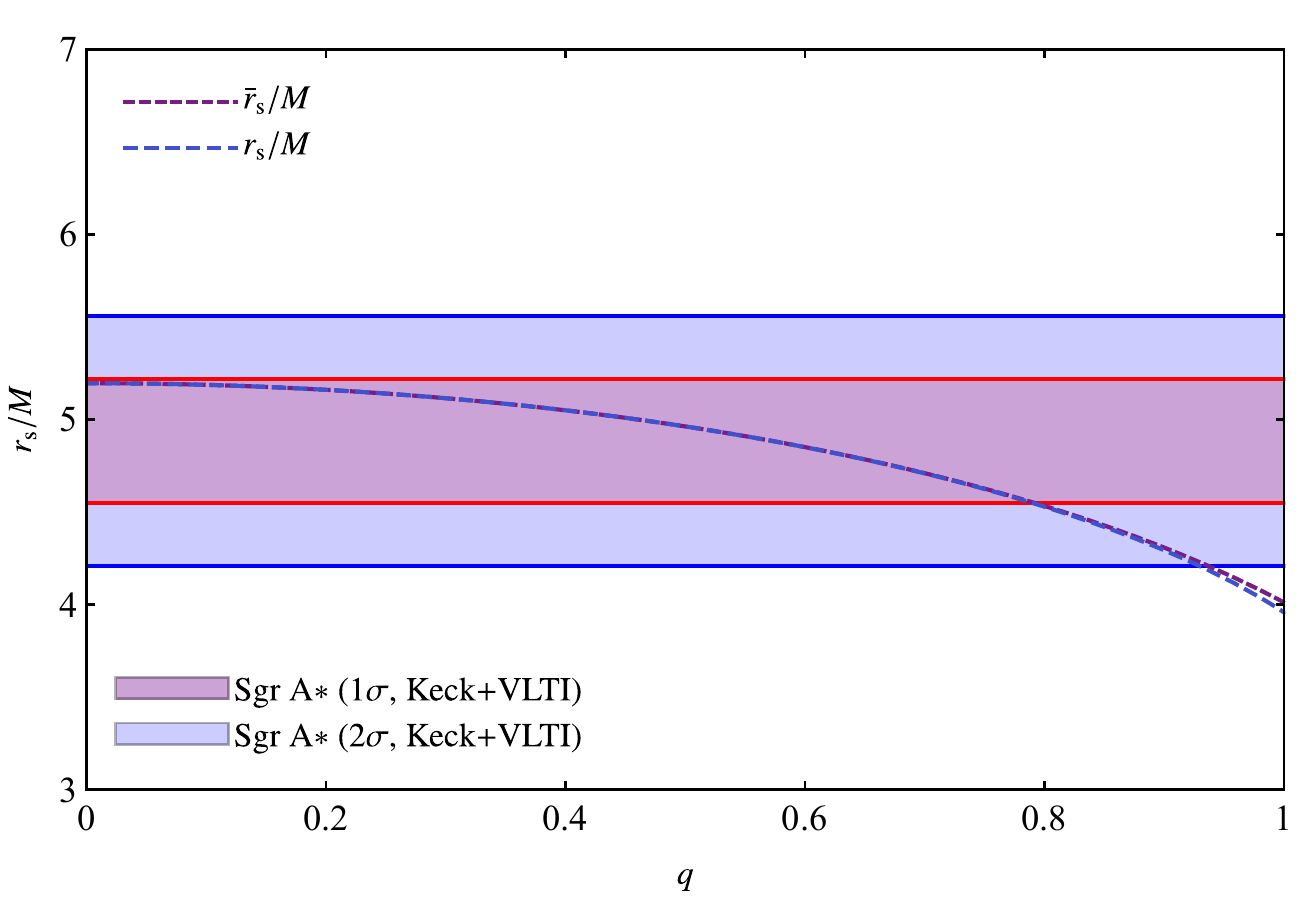}
		\caption{\footnotesize{Shadow radius of fully RBHs with $h = 2$, considering the standard and effective geometries, as a function of $q$. The light blue and purple regions are consistent with the EHT horizon-scale image of Sgr A$^{\star}$ at $1\sigma$ and $2\sigma$ constraints, respectively. The white regions denote the excluded regions.}}
		\label{sgr}
	\end{centering}
\end{figure}

In Fig.~\ref{sgr}, we compare our results for the shadow radius of the case $h = 2$ with the Keck/VLTI observational data for Sgr A$^{\star}$. We note that the data set the $1\sigma$ upper limits $Q \lesssim 0.7999 M$ and $Q \lesssim 0.79979 M$, considering the effective and standard geometries, respectively. On the other hand, the $2\sigma$ constraints lead to the upper limits $Q \lesssim 0.9466M$ and $Q \lesssim 0.9392M$, considering the standard and effective geometries, respectively. Therefore, observational data rule out the possibility that Sgr A$^{\star}$ is an extremally charged fully RBH with $h = 2$, as occurs for the RN case, up to $2\sigma$ constraints. Yet, moderately charged fully RBHs with $h = 2$ are still feasible.

\begin{figure}[!htbp]
	\begin{centering}
		\includegraphics[width=0.6\columnwidth]{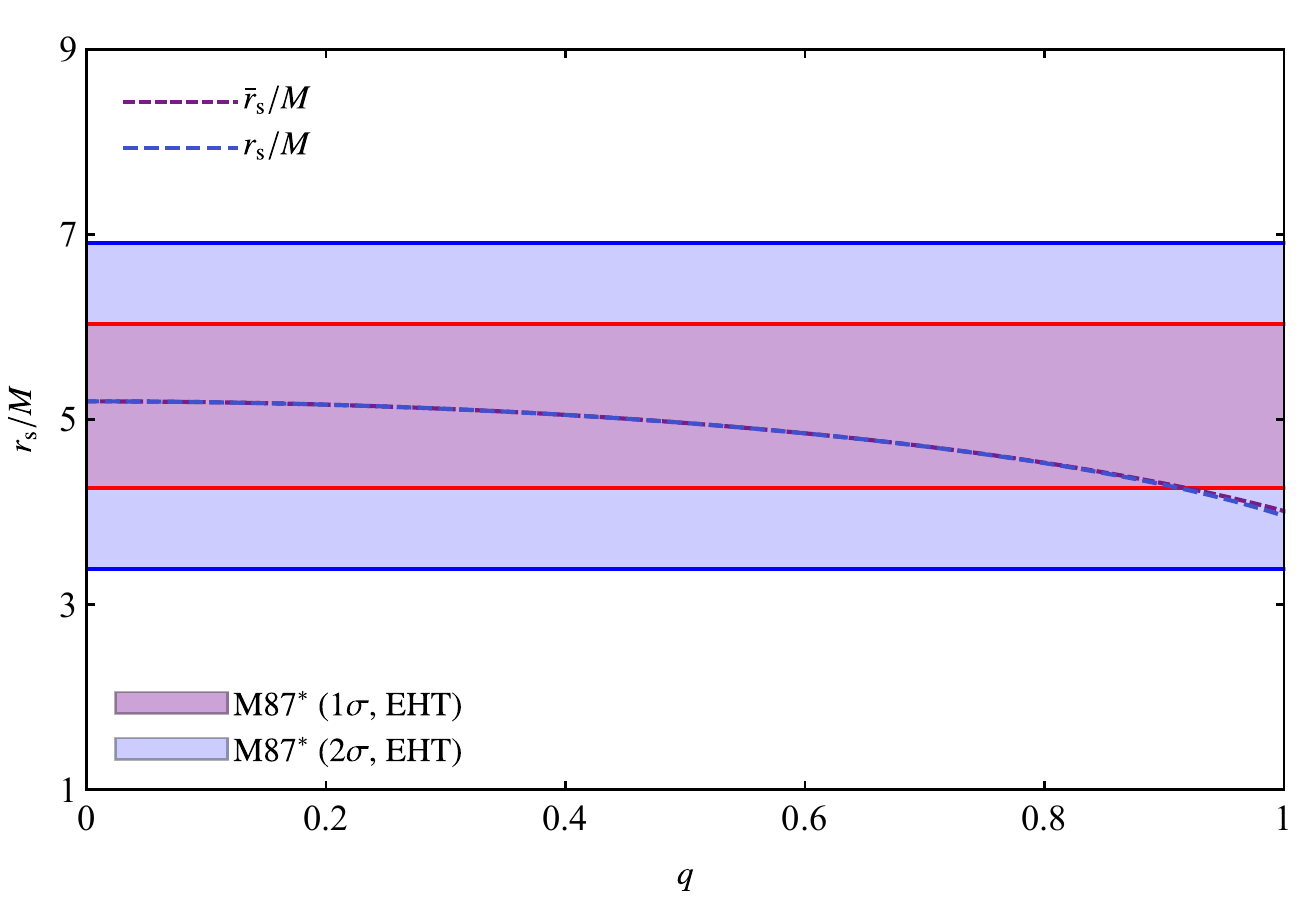}
		\caption{\footnotesize{Shadow radius of fully RBHs with $h = 2$, considering the standard and effective geometries, as a function of $q$. The light blue and purple regions are consistent with the EHT horizon-scale image of M87$^{\star}$ at $1\sigma$ and $2\sigma$ constraints, respectively. The white regions denote the excluded regions.}}
		\label{m87}
	\end{centering}
\end{figure}

In Fig.~\ref{m87}, we compare our results for the shadow radius of the case $h = 2$ with the EHT observational data for M87$^{\star}$. We note that the EHT observations for M87$^{\star}$ set the $1\sigma$ upper limits $Q \lesssim 0.9282 M$ and $Q \lesssim 0.9219 M$  for the effective and standard geometries, respectively. On the other hand, the results for $Q \leq Q_{\text{ext}}$ are consistent with the $2\sigma$ constraints. Consequently, the EHT observations show that M87$^{\star}$ is consistent with extremally charged fully RBH solutions with $h = 2$ for $2\sigma$ constraints, but, when considering $1\sigma$ constraints, only for at most moderately charged fully RBHs with $h = 2$.

The bounds for the charge-to-mass ratio for the case $h = 0$ are very similar to those of the case $h = 2$. Consequently, the conclusions are basically the same. To illustrate this, in Tab.~\ref{tabh0}, we summarize the main results for the $h = 0$ case.
\begin{table}[hbtp!]
	\centering
	\begin{tabular}{||c|c|c|c|c||} 
		\hline
		$\text{BHs}$ & \multicolumn{2}{|c|}{Sgr A$^{\star}$} & \multicolumn{2}{|c||}{M87$^{\star}$} \\ \cline{1-5}
		$\text{Constraints}$ & $1\sigma$ & $2\sigma$ & $1\sigma$ & $2\sigma$    \\ \hline
		$\left(Q/M\right)^{\text{SG}}$ &$0.797959$ & $0.938977$ & $0.921972$ &  --   \\  \hline
		$\left(Q/M\right)^{\text{EG}}$ & $0.797963$ & $0.936068$ & $0.928275$ & -- \\  \hline
	\end{tabular}
	
	\caption{\footnotesize{ Bounds to the charge-to-mass ratio of fully RBHs with $h = 0$, considering the standard and effective geometries and the observational data for Sgr A$^{\star}$ and M87$^{\star}$. The symbol ``--'' denotes that the BHs for $h = 0$ are within the range of the data constraints, regardless of the value of $Q/M$.}}
	\label{tabh0}
\end{table} 

It is also interesting to verify how large the charge-to-mass ratio $Q/M$ is in physical units, considering the EG, and to compare it to some bounds obtained from astrophysical considerations. To do this, we first restore the physical units. Consider $Q = \rho M$, where $\rho$ is a numerical value. In physical units, $M \rightarrow G M/c^{2}$ and $Q^{2} \rightarrow G Q^{2}/4\pi \epsilon_{0} c^{4}$, where $G$ is the gravitational constant, $c$ is the speed of light, and $\epsilon_{0}$ is the vacuum permittivity. Therefore, we have
\begin{equation}
	\label{physicalcharge}Q = 2 M \rho \sqrt{G \pi \epsilon_{0}}.
\end{equation}

As the results for $h = 0$ are similar to those of $h = 2$, we consider the values of the ratio $Q/M$ for case $h = 2$.  In this case, for constraints $1\sigma$ and $2\sigma$, considering Sgr A$^{\star}$, we have $Q \approx 5.41716\times10^{26}C$ and $Q \approx 6.41065\times10^{26}C$, respectively, where we use $M_{\text{Sgr} A^{\star}} = 3.951\times10^{6}M_\odot$~\cite{Do:2019txf}. Concerning M87$^{\star}$, we have $Q \approx 1.03421\times10^{30}C$ and $Q \approx 1.12621\times10^{30}C$, where we use $M_{\text{M87}^{\star}} = 6.5\times10^{9}M_\odot$~\cite{EventHorizonTelescope:2019ggy}, at the $1\sigma$ and $2\sigma$ levels, respectively.

The results for the charge-to-mass ratio obtained using the data for Sgr A$^{\star}$ can be compared with the charge constraints of Sgr A$^{\star}$ obtained in Refs.~\cite{Zajacek:2018vsj,Zajacek:2018ycb}. The extremal charge of Sgr A$^{\star}$ is $Q^{\text{ext}}_{\text{Sgr} A^{\star}} \approx 4 \times 10^{27}C$~\cite{Iorio:2012dbo}. However, in astrophysical scenarios, the charge of Sgr A$^{\star}$ is expected to have the upper limits $Q \lesssim 3 \times 10^{8} C$~\cite{Zajacek:2018vsj} or $Q \lesssim 3 \times 10^{15} C$~\cite{Zajacek:2018ycb}, depending on the astrophysical setup. Therefore, our results for fully RBH solutions with $h = 0$ and $h = 2$ show that their charges are close to the extremally charged case of Sgr A$^{\star}$, but remarkably larger than the upper limits on the Sgr A$^{\star}$ charge. We emphasize that these charge values for the fully RBHs represent theoretical upper limits derived from shadow geometry constraints, rather than expected physical charges, as real astrophysical environments would neutralize such magnitudes. Nonetheless, mapping these theoretical bounds is a critical step in understanding the parameter space of the h-family.

We also point out that lacking exact rotating counterparts of the static fully RBH solutions, we have ignored the effects of rotation of Sgr A$^{\star}$ and M87$^{\star}$ on the value of $Q/M$. However, constraining the charge parameter using static models serves as a fundamental {\textquoteleft null hypothesis\textquoteright} test and is a standard practice in the literature. This analysis allows us to isolate the specific phenomenological signatures of the electric charge, providing a baseline for future rotating generalizations. Moreover, this method is an effective mechanism for verifying how physical the BH charge is. Slowly rotating, if not exact, solutions could be derived by known procedures~\cite{Azreg2014}, which will be the subject of future investigations.

%%%%%%%%%%%%%%%%%%%%%%%%%%%%%%%%%%%%%%%%%%%%%%%%%%
\section{Gravitational and kinematic redshifts\label{sech2red}}
%%%%%%%%%%%%%%%%%%%%%%%%%%%%%%%%%%%%%%%%%%%%%%%%%%

%%%%%%%%%%%%%%%%%%%%%%%%%%%%%%%%%%%%%%%%%%%%%%%%%%
\subsection{Main equations}
%%%%%%%%%%%%%%%%%%%%%%%%%%%%%%%%%%%%%%%%%%%%%%%%%%

We assume the presence of a star orbiting a NED RBH given by a general metric of the form given by Eq.~\eqref{sss0}, with $A(r) = g_{tt}(r)$, $B(r) = g_{rr}(r)$, and $C(r) = g_{\varphi \varphi}(r)$\footnote{In what follows, we assume the motion in the equatorial plane, thus $r^{2}\dd \Omega^{2} \rightarrow r^{2}\dd\varphi^{2}$. Therefore, for simplicity, we label $C(r) = g_{\varphi \varphi}(r)$.}. The star emits radiation, the frequency of which is measured by a detector (usually at spatial infinity). We aim to determine the maximum and minimum values of the redshift (we use the term redshift to mean either redshift or blueshift). In LED, such expressions are known in the literature in the case where $g_{\varphi \varphi}(r)=r^2$~\cite{Lake:2003tr}, considering the metric signature ($-,+,+,+$) and $\theta = \pi/2$. We will derive the most general expression of the redshift that applies to NED as well. While we are using an SSS metric with signature ($+,-,-,-$), we will provide an expression for the redshift and intermediate expressions that are valid for both SSS-metric signatures ($+,-,-,-$) and ($-,+,+,+$). We introduce the symbol $\epsilon$, given by $\epsilon =+1$ if the metric signature is ($+,-,-,-$), and $\epsilon =-1$, if the metric signature is ($-,+,+,+$). We have $\epsilon g_{tt}=|g_{tt}|$ and $-\epsilon g_{\varphi\varphi}=|g_{\varphi\varphi}|$.

The star, emitting radiation, is supposed to move in a circular orbit in the plane $\theta=\pi/2$ with 4-velocity $u^\mu =(u^t,0,0,u^\varphi)$. Circular motion is treated in many references, and we can show that the specific energy $E$ and the angular momentum $L$ of the star, treated as a particle following a circular geodesic, are given by
\begin{align}
	\label{s1}&E^2=\frac{|g_{tt}|\,(\ln|g_{\varphi\varphi}|)^{\prime}}{(\ln|g_{\varphi\varphi}|)^{\prime}-(\ln|g_{tt}|)^{\prime}}\,,\\
	\label{s2}&L^2=\frac{|g_{\varphi\varphi}|\,(\ln|g_{tt}|)^{\prime}}{(\ln|g_{\varphi\varphi}|)^{\prime}-(\ln|g_{tt}|)^{\prime}}\,.
\end{align}
They satisfy, respectively, the conservation equations:
\begin{equation}\label{s3}
	E=\epsilon g_{tt}u^t=|g_{tt}|u^t\,,\quad L=-\epsilon g_{\varphi\varphi}u^\varphi=|g_{\varphi\varphi}|u^\varphi\,.
\end{equation}

If $k^\mu$ is the 4-momentum of the photon emitted by the star, the frequency of the emitted photon is $\omega_e=\epsilon \bar{g}_{\mu\nu}k_e^\mu u^\nu$, where $u^\mu =(u^t,0,0,u^\varphi)$ is the 4-velocity of the star and $k_e^\mu$ is the value of $k^\mu$ at the star's position. The frequency of the photon absorbed (received) by the detector is $\omega_d=\epsilon \bar{g}_{\mu\nu}(\infty)k_d^\mu U^\nu$, where $U^\mu =(1/\sqrt{|g_{tt}(\infty)|},0,0,0)$ is the 4-velocity of the detector and $k_d^\mu=k^\mu(\infty)$. The extreme values of the redshift are obtained when the photon ray, in the plane $\theta=\pi/2$, is tangent to the circle, i.e., when $k^\mu=(k^t,0,0,k^\varphi)$. Now, the EG $\bar{g}_{\mu\nu}$~\eqref{eg1} is of the form $\bar{g}_{\mu\nu}=\alpha_{\mu\nu}g_{\mu\nu}$ [no summation and $\text{sgn}(\bar{g}_{\mu\nu})=\text{sgn}(g_{\mu\nu})$], where $\alpha_{\mu\nu}$ are scalar (positive) functions of coordinates given in terms of the NED fields. Using $\bar{g}_{\mu\nu}=\alpha_{\mu\nu}g_{\mu\nu}$ along with~\eqref{s3}, we obtain that the frequencies are given by
\begin{equation}\label{s4}
	\omega_e=\alpha_{tt}Ek_e^t - \alpha_{\varphi\varphi}Lk_e^\varphi\,,\qquad
	\omega_d=k_d^t\sqrt{\alpha_{tt}(\infty)|g_{tt}(\infty)|}\,.
\end{equation}

The redshift $z$ is defined by $z\equiv(\lambda_d - \lambda_e)/\lambda_e$~\cite{schutz2022first}, where ($\lambda_e,\lambda_d$) are the corresponding wavelengths. In terms of ($\omega_e,\omega_d$), the redshift can be written as
\begin{equation}\label{s5}
	z=\frac{\omega_e}{\omega_d}-1=\frac{k_e^t}{k_d^t\sqrt{\alpha_{tt}(\infty)|g_{tt}(\infty)|}}\Big(\alpha_{tt}E-\alpha_{\varphi\varphi}L\frac{k_e^\varphi}{k_e^t}\Big)-1\,,
\end{equation}
where $E$ and $L$ are given by Eqs.~\eqref{s1} and \eqref{s2}, respectively. To evaluate $k_e^\varphi/k_e^t$, we use the fact that the photon 4-momentum satisfies the constraint $\bar{g}_{\mu\nu}k^\mu k^\nu=0$, which yields
\begin{equation}\label{s6}
	\frac{k_e^\varphi}{k_e^t}=\pm \sqrt{\frac{|\bar{g}_{tt}|}{|\bar{g}_{\varphi\varphi}|}}=\pm \sqrt{\frac{\alpha_{tt}|g_{tt}|}{\alpha_{\varphi\varphi}|g_{\varphi\varphi}|}}\,.
\end{equation}
Without loss of generality, we can always choose $L>0$. With this choice at hand, the ``$+$'' sign corresponds to $k_e^\varphi>0$, which means that at the moment of emission the star was approaching the detector (a blueshift effect), and this yields a minimum value for $z$, $z_\text{min}=z_+$. The ``$-$'' sign corresponds to $k_e^\varphi<0$, meaning the star was receding at the moment of emission (a redshift effect), yielding a maximum value for $z$, $z_\text{max}=z_-$. This reduces the expressions of $z_\pm$ to
\begin{equation}\label{s7}
	z_\pm=\frac{k_e^t}{k_d^t\sqrt{\alpha_{tt}(\infty)|g_{tt}(\infty)|}}\bigg[\alpha_{tt}E \mp \alpha_{\varphi\varphi}L\sqrt{\frac{\alpha_{tt}|g_{tt}|}{\alpha_{\varphi\varphi}|g_{\varphi\varphi}|}}\,\bigg]-1\,,
\end{equation}
where the upper (lower) sign on the left-hand side corresponds to the upper (lower) sign on the right-hand side. The photon obeys conservation equations similar to Eq.~\eqref{s3}, but in the EG, where its path is a geodesic, we have $\bar{E}_{\text{ph}}=\epsilon \bar{g}_{tt}k^t=|\bar{g}_{tt}|k^t$ [compare with~\eqref{eqm1_EG} where the constants $\bar{E}_{\text{ph}}$ and $E_{\text{ph}}$ are proportional]. This yields $|\bar{g}_{tt}|k_e^t=|\bar{g}_{tt}(\infty)|k_d^t$, i.e.,
\begin{equation}\label{s8}
	\frac{k_e^t}{k_d^t}=\frac{|\bar{g}_{tt}(\infty)|}{|\bar{g}_{tt}|}=\frac{\alpha_{tt}(\infty)|g_{tt}(\infty)|}{\alpha_{tt}|g_{tt}|}\,,
\end{equation}
and
\begin{equation}\label{s9}
	z_\pm=\frac{\sqrt{\alpha_{tt}(\infty)|g_{tt}(\infty)|}}{\alpha_{tt}|g_{tt}|}\bigg[\alpha_{tt}E \mp \alpha_{\varphi\varphi}L\sqrt{\frac{\alpha_{tt}|g_{tt}|}{\alpha_{\varphi\varphi}|g_{\varphi\varphi}|}}\,\bigg]-1\,.
\end{equation}
Finally, substituting the expressions for $E$ and $L$, given by Eqs.~\eqref{s1}-\eqref{s2}, we obtain the general expression of the redshift for an SSS metric in the presence of a NED field, namely
\begin{multline}\label{s10}
	z_\pm=\frac{\alpha_{tt}(\infty)\sqrt{|g_{tt}(\infty)|}}{\alpha_{tt}\sqrt{|g_{tt}|}\sqrt{(\ln|g_{\varphi\varphi}|)'-(\ln|g_{tt}|)'}}\bigg(\sqrt{(\ln|g_{\varphi\varphi}|)'} \mp \sqrt{(\ln|g_{tt}|)'}\sqrt{\frac{\alpha_{tt}}{\alpha_{\varphi\varphi}}}\,\bigg)-1\,,
\end{multline}
\begin{equation}\label{s10}
	z_\pm=\sqrt{\frac{\alpha_{tt}(\infty)|g_{tt}(\infty)|}{\alpha_{tt}|g_{tt}|}}~\frac{\sqrt{\alpha_{tt}(\ln|g_{\varphi\varphi}|)'} \mp \sqrt{\alpha_{\varphi\varphi}(\ln|g_{tt}|)'}}{\sqrt{(\ln|g_{\varphi\varphi}|)'-(\ln|g_{tt}|)'}}-1\,,	
\end{equation}
where $z_\text{min}=z_+$ and $z_\text{max}=z_-$. In LED, the functions $\alpha_{\mu\nu}$ are all unity. In our case, as previously noticed, $\alpha_{tt}(\infty)=1$ and $|g_{tt}(\infty)|=1$ because Eq.~\eqref{eg1} is asymptotically flat.

Upon applying Eq.~\eqref{s10} to our case under investigation, with $\alpha_{tt}=1/\Phi$, $\alpha_{\varphi\varphi}=1/\mathcal{L}_F$, $|g_{tt}|=f(r)$~\eqref{n84a2}, and $|g_{\varphi\varphi}|=r^2$, we obtain,
\begin{equation}\label{s11}
	z_\pm=\frac{1}{\sqrt{f-\dfrac{rf'}{2}}}\Bigg[1 \mp \sqrt{\frac{rf'}{2f}}\sqrt{\frac{\Phi}{\mathcal{L}_F}}\,\Bigg]-1\,.
\end{equation}
Recall that the condition for massless particles in a circular orbit is given by $2f_{c}-r_{c}f^{\prime}_{c} = 0$ [cf. Eq.~\eqref{bcrc}]. Therefore, Eq.~\eqref{s11} shows that the kinematic redshift becomes extremely large as $2f-rf^{\prime} \rightarrow 0$, and the emitter cannot orbit at radii $r_e \le r_c$, which implies that this region is physically inaccessible for massive emitting stars.

If the star were not moving ($u^\varphi =0$ and $L=0$), see Eq.~\eqref{s3}, we would obtain the purely gravitational redshift. The star's 4-velocity would be $u^\mu =(1/\sqrt{|g_{tt}|},0,0,0)$, the emitted frequency $\omega_e=\alpha_{tt}k_e^t\sqrt{|g_{tt}|}$, and
\begin{equation}\label{s12}
	z_{\text{grav}}=\frac{\alpha_{tt}k_e^t\sqrt{|g_{tt}|}}{k_d^t\sqrt{\alpha_{tt}(\infty)|g_{tt}(\infty)|}}-1\,,
\end{equation}
where $k_e^t/k_d^t$ is still given by Eq.~\eqref{s8}. Finally,
\begin{equation}\label{s13}
	z_{\text{grav}}=\frac{\sqrt{\alpha_{tt}(\infty)|g_{tt}(\infty)|}}{\sqrt{|g_{tt}|}}-1\,.
\end{equation}
The kinematic redshift (Doppler part)\footnote{The wavelength received by the detector is $\lambda_d=\lambda_e+\delta\lambda_{\text{grav}}+\delta\lambda_{\text{kin}}$ where $\delta\lambda_{\text{grav}}$ is the variation in wavelength due to gravity and $\delta\lambda_{\text{kin}}$ is the corresponding kinematic variation. The total redshift is $z=[(\lambda_e+\delta\lambda_{\text{grav}}+\delta\lambda_{\text{kin}})-\lambda_e]/\lambda_e$ yielding $z=(\delta\lambda_{\text{grav}}/\lambda_e)+(\delta\lambda_{\text{kin}}/\lambda_e)=z_{\text{grav}}+z_{\text{kin}}$.}, i.e., $z_{\pm (\text{kin})}=z_\pm - z_{\text{grav}}$, can be written as
\begin{align}\label{s14}
	z_{\pm (\text{kin})}=\sqrt{\frac{\alpha_{tt}(\infty)|g_{tt}(\infty)|}{\alpha_{tt}|g_{tt}|}}~\Bigg[\frac{\sqrt{\alpha_{tt}(\ln|g_{\varphi\varphi}|)'} \mp \sqrt{\alpha_{\varphi\varphi}(\ln|g_{tt}|)'}}{\sqrt{(\ln|g_{\varphi\varphi}|)'-(\ln|g_{tt}|)'}}-\sqrt{\alpha_{tt}}\Bigg]\,.
\end{align}
This is the general expression for the kinematic redshift. For our case, we obtain
\begin{equation}\label{s15}
	z_{\pm (\text{kin})}=\frac{1}{\sqrt{f-\dfrac{rf'}{2}}}
	\Bigg[1 \mp \sqrt{\frac{rf'}{2f}}\sqrt{\frac{\Phi}{\mathcal{L}_F}}-\sqrt{1-\dfrac{rf'}{2f}}\,\Bigg]\,.
\end{equation}
By~\eqref{4b} we see that for all values of $h$, the only way to re-obtain the results of LED is when $r$ goes to infinity. In this limit, $1-rf'/(2f)\to 1$, $\Phi\to \mathcal{L}_F\to 1$, and the last equation reduces to the symmetric expression
\begin{equation}
	z_{\pm (\text{kin})}\simeq \mp \frac{1}{f}\sqrt{\dfrac{rf'}{2}} \,.    
\end{equation}
In the same limit $r\to\infty$, the symmetric equation obtained from~\eqref{s14} is
\begin{equation}
	z_{\pm (\text{kin})}\simeq \mp \frac{1}{g_{tt}}\sqrt{\dfrac{(g_{tt})'}{(\ln|g_{\varphi\varphi}|)'}}\,.
\end{equation}
Such an observation of symmetry in LED, $z_{+ (\text{kin})}=-z_{- (\text{kin})}$ ($z_{\text{max (kin)}}=-z_{\text{min (kin)}}$), was already noticed in~\cite{Mass-spin22} for the case of the Kerr BH.

%%%%%%%%%%%%%%%%%%%%%%%%%%%%%%%%%%%%%%%%%%%%%%%%%%
\subsection{Results for $h =0$ and $h = 2$ cases}
%%%%%%%%%%%%%%%%%%%%%%%%%%%%%%%%%%%%%%%%%%%%%%%%%%

\begin{figure}[!htbp]
	\begin{centering}
		\includegraphics[width=0.47\columnwidth]{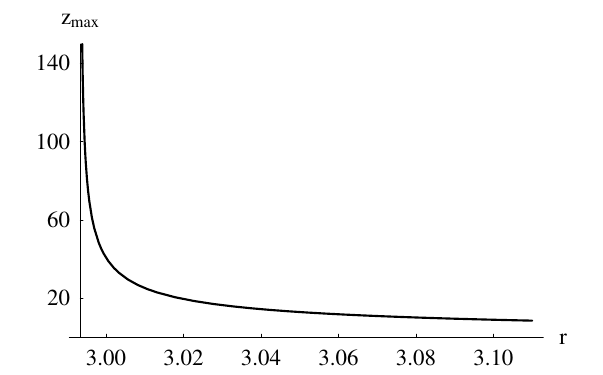}
		\includegraphics[width=0.47\columnwidth]{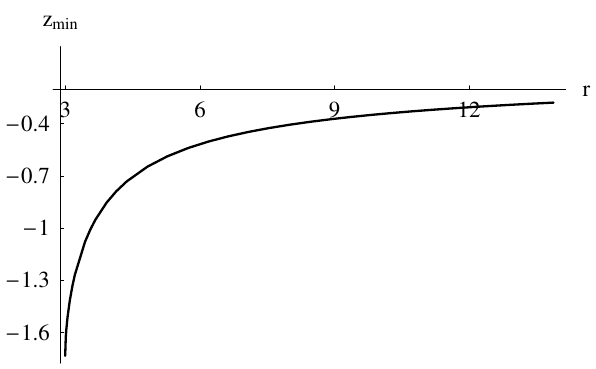}
		\caption{\footnotesize{Plots of the kinematic redshifts $z_{\text{max (kin)}}$ (left panel -- redshift) and $z_{\text{min (kin)}}$ (right panel -- blueshift) for the NED metric~\eqref{n84a2}. We took $M=1$ and $Q=0.1$.}}
		\label{figmmh2a}
	\end{centering}
\end{figure}

\begin{figure}[!htbp]
	\begin{centering}
		\includegraphics[width=0.47\columnwidth]{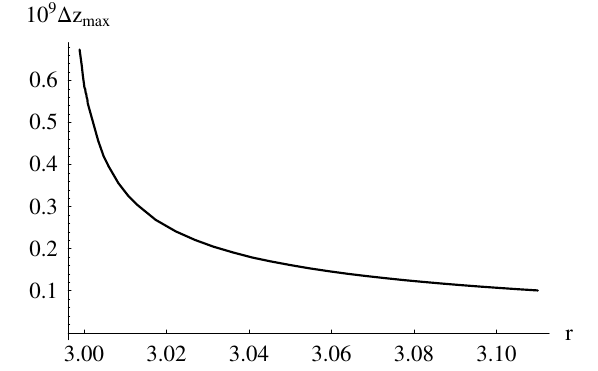}
		\includegraphics[width=0.47\columnwidth]{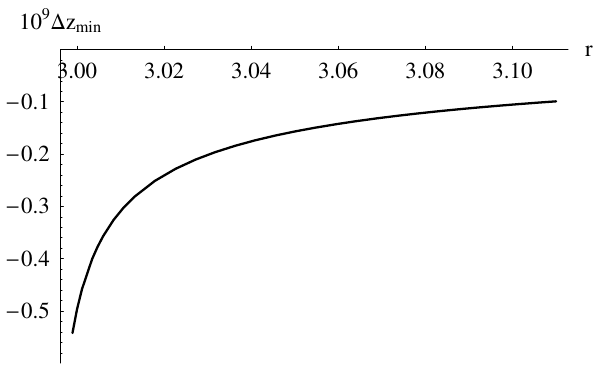}
		\caption{\footnotesize{Plots of differences of kinematic redshifts times $10^9$: $10^9\Delta z_{\text{max}}$ (left panel) and $10^9\Delta z_{\text{min}}$ (right panel), where $\Delta z_{\text{max}}=z_{\text{max (kin)}}-z_{\text{max (kin-RN)}}$ and $\Delta z_{\text{min}}=z_{\text{min (kin)}}-z_{\text{min (kin-RN)}}$ with $z_{\text{max (kin)}}$ and $z_{\text{min (kin)}}$ being the kinematic redshifts related to the NED metric~\eqref{n84a2} and $z_{\text{max (kin-RN)}}$ and $z_{\text{min (kin-RN)}}$ being the kinematic redshifts for the RN BH. We took $M=1$ and $Q=0.1$.}}
		\label{figmmh2b}
	\end{centering}
\end{figure}

In Fig.~\ref{figmmh2a}, we plot the kinematic redshifts $z_{\text{max (kin)}}$ and $z_{\text{min (kin)}}$ for the NED metric~\eqref{n84a2}, and in Fig.~\ref{figmmh2b}, we plot the difference between the kinematic redshifts for the NED metric~\eqref{n84a2} and the corresponding entities for the RN BH. The second plot shows that the maximum kinematic (redshift) for the NED metric~\eqref{n84a2} is greater than that for the RN BH and that the minimum kinematic (blueshift) for the NED metric~\eqref{n84a2} is smaller than that for the RN BH. This difference is more pronounced in the vicinity of the smallest circular geodesic defined by $2f-rf'=0$.

Despite the difference between the NED metrics~\eqref{n84a} and~\eqref{n84a2}, their kinematic redshifts for $M=1,\,Q=0.1$ are almost identical. In Table~\ref{tab1}, we provide values for $z_{\text{max (kin)}}$ and $z_{\text{min (kin)}}$ for both metrics for different values of the radial coordinate $r$ when $M=1,\,Q=0.1$. Note that the corresponding values of $x\,(=r/K)$ are different due to the different values of $K$ in~\eqref{h2K} and in~\eqref{h0K} for a given set of $M,\,Q$ values. For these values of $M=1,\,Q=0.1$, $x$ is relatively large [$r=2.998\leftrightarrow x=438.231$~\eqref{n84a} and $r=2.998\leftrightarrow x=539.83$~\eqref{n84a2}] and this explains why the kinematic redshifts are almost equal and equal to each other and to the corresponding values of the RN BH.

%\begin{table}[!t]
\begin{table}[h]
	\begin{centering}
		\centering
		\begin{tabular}{||c|c|c|c|c|c||}
			\hline 
			$r$ & $r_{\text{p}}$ & $z_{\text{min (kin)}}$~\eqref{n84a} & $z_{\text{min (kin)}}$~\eqref{n84a2} & $z_{\text{max (kin)}}$~\eqref{n84a} & $z_{\text{max (kin)}}$~\eqref{n84a2} \\
			\hline 
			\hline 
			2.998 & 2.998 & $-1.67116$ & $-1.67116$ & 48.8787 & 48.8787 \\
			3.010 & 3.010 &  $-1.61259$ & $-1.61259$ & 25.061 & 25.061 \\
			3.110 & 3.110 &  $-1.3937$ & $-1.3937$ & 8.38739 & 8.38739 \\
			3.200 & 3.200 &  $-1.27739$ & $-1.27739$ & 5.8935 & 5.8935 \\
			5.100 & 5.100 &  $-0.608431$ & $-0.608431$ & 1.15783 & 1.15783 \\
			\hline 
		\end{tabular}
	\end{centering}
	\caption{\label{tab1}{\footnotesize Values of $z_{\text{max (kin)}}$ and $z_{\text{min (kin)}}$ corresponding to the NED metrics~\eqref{n84a} and~\eqref{n84a2}. We took $M=1,\,Q=0.1$. In this table and in Table~\ref{tab2}, the radial coordinate $r_{\text{p}}\equiv \sqrt{\bar{g}_{\varphi\varphi}}\big|_{\theta=\frac{\pi}{2}} =r/\sqrt{\mathcal{L}_F}$ is the perimetral radius, which is a meaningful geometrical quantity to compare distances in two different geometries (as done in~\cite{ dePaula:2023ozi}). The values of $r_{\text{p}}$ are the same for both metrics~\eqref{n84a} and~\eqref{n84a2}.}}
\end{table}

\begin{table}[h]
	\begin{centering}
		\centering
		\begin{tabular}{||c|c|c|c|c|c|c||}
			\hline 
			$r$ & $r_{\text{p}}~\eqref{n84a}$ & $r_{\text{p}}~\eqref{n84a2}$ & $z_{\text{min (kin)}}$~\eqref{n84a} & $z_{\text{min (kin)}}$~\eqref{n84a2} & $z_{\text{max (kin)}}$~\eqref{n84a} & $z_{\text{max (kin)}}$~\eqref{n84a2} \\
			\hline 
			\hline
			2.501 & 2.501 & 2.5001 & $-1.81626$ & $-2.27198$ & 351.067 & 1217.65 \\
			2.511 & 2.511 & 2.5101 & $-1.72415$ & $-1.72878$ & 33.3171 & 56.5762 \\ 
			2.998 & 2.998 & 2.9975 & $-1.06611$ & $-1.06603$ & 3.29278 & 3.32943 \\
			3.110 & 3.110 & 3.1095 & $-1.00015$ & $-1.00001$ & 2.85167 & 2.87895 \\
			5.100 & 5.100 & 5.0999 & $-0.552904$ & $-0.552797$ & 0.986436 & 0.990024 \\
			\hline 
		\end{tabular}
	\end{centering}
	\caption{\label{tab2}{\footnotesize Values of $z_{\text{max (kin)}}$ and $z_{\text{min (kin)}}$ corresponding to the NED metrics~\eqref{n84a} and~\eqref{n84a2}. We took $M=1,\,Q=0.79$. The values of $r_{\text{p}}$ are nearly the same for the metrics~\eqref{n84a} and~\eqref{n84a2}.}}
\end{table}

We can have a better idea if we go through numerical estimations. For the average value $\lambda_e=545$ nm of the visible color spectrum (from 380 nm to 710 nm) and $r=3.0 M$, we obtain, using the kinematic redshift expression~\eqref{s15} and the metric~\eqref{n84a2} taking $M=1$ and $Q=0.1$, $\lambda_{d\text{ (max)}}-\lambda_{d\text{ (max RN)}}=3.2453\times 10^{-7}$ nm, which might be difficult to measure. However, if the light emitted passes adjacent to the smallest circular geodesic defined by $2f-rf'=0$, where $r\simeq 2.99338 M$ (not very different from $r=3.0 M$), we obtain, for the same average value $\lambda_e$, $\lambda_{d\text{ (max)}}-\lambda_{d\text{ (max RN)}}=0.0000619713$ nm, which represents more than 190 times the previous value. It is worth noting that near-boundary examples do not correspond to stable stellar orbits, as they lie well within the innermost stable circular orbit (ISCO). However, this regime is relevant for the plunging region of the BH, namely, the region between the event horizon and the ISCO, since redshifts in this zone are important when one considers accretion disks~\cite{2019ApJ87412S,2019ApJ887145S,Zhu:2012vf}.

For much larger values of $Q/M$, say $M=1,\,Q=0.79$, $x$ is on the order of $r$ [$r=2.998\leftrightarrow x=7.0218$~\eqref{n84a} and $r=2.998\leftrightarrow x=8.64973$~\eqref{n84a2}], and the kinematic redshifts tend to diverge, as shown in Table~\ref{tab2}.

For stars orbiting in the vicinity of the smallest circular geodesic, defined by $2f-rf'=0$, it is much easier to distinguish, using the kinematic redshift, NED metrics from each other and from the RN BH in the case of a moderate ratio $Q/M\gtrsim 0.79$.

As we can see from Eqs.~\eqref{n86} and~\eqref{n862}, the effect of including the NED fields was primarily to remove the singularity at the center $r=0$, without greatly modifying the physical and geometric properties of the RN BH. This character is also observed in the kinematic redshift, where, for instance, the \emph{first eight} terms of the power expansion of $z_{- (\text{kin})}$, for the metric~\eqref{n84a2}, are identical to those of the maximum kinematic redshift of the RN BH, i.e.,
\begin{multline}
	z_{\text{max (kin)}}-z_{\text{max (kin-RN)}}=\frac{4\times 2^{1/4} M}{\sqrt{\pi } |Q|} \Big(\frac{1}{x}\Big)^{9/2}+\frac{4\times 2^{3/4} M \Big(50 M^2-7 Q^2\Big)}{5
		\pi ^{3/2} |Q|^3} \Big(\frac{1}{x}\Big)^{11/2}\\+\frac{48 M^2}{5 \pi ^2 Q^2} \Big(\frac{1}{x}\Big)^6+\mathcal{O}\Big(\frac{1}{x}\Big)^{13/2}\,,    
\end{multline}
where $x=r/K$~\eqref{h2K}. There is a similar expression for $z_{\text{min (kin)}}-z_{\text{min (kin-RN)}}$:
\begin{multline}
	z_{\text{min (kin)}}-z_{\text{min (kin-RN)}}=-\frac{4\times 2^{1/4} M}{\sqrt{\pi } |Q|} \Big(\frac{1}{x}\Big)^{9/2}-\frac{4\times 2^{3/4} M \Big(50 M^2-7 Q^2\Big)}{5
		\pi ^{3/2} |Q|^3} \Big(\frac{1}{x}\Big)^{11/2}\\+\frac{48 M^2}{5 \pi ^2 Q^2} \Big(\frac{1}{x}\Big)^6+\mathcal{O}\Big(\frac{1}{x}\Big)^{13/2}\,.   
\end{multline}

%%%%%%%%%%%%%%%%%%%%%%%%%%%%%%%%%%%%%%%%%%%%%%%%%%
\section{Final Remarks}\label{sec:remarks}
%%%%%%%%%%%%%%%%%%%%%%%%%%%%%%%%%%%%%%%%%%%%%%%%%%

NED-sourced spacetimes are interesting testing grounds for investigating the properties of RBHs and also the imprints of very strong electromagnetic fields in compact object physics. This is supported by two perspectives. First, NED provides a medicine to cure curvature singularities, making a rich class of RBH solutions feasible. Second, photons in NED follow null geodesics according to an effective light cone. To broaden our knowledge on these topics, we have addressed some of the physical and geometrical properties of a recent family of electrically charged RBHs --- the so-called h-family~\cite{Azreg-Ainou:2025tuj}. In particular, we explored the motion of photons, considering the EG, by investigating the shadow radius, the weak deflection angle, and the gravitational and kinematic redshifts.

Considering the shadow radius, we have derived the correct formula for a static observer placed at a given radial position sending a light ray to the past with an angle $\beta$ with respect to the radial direction. Our findings show that, for the h-family of BHs studied here, the EG does not significantly affect the shadow radius. We also imposed some astrophysical constraints on the BH charge-to-mass ratio $Q/M$ based on observational data. In particular, we have shown that the BH charges in SI units are close to the extremally charged case of Sgr A$^{\star}$, but substantially higher than the upper limits on the Sgr A$^{\star}$ charge.

One of our priorities in this work was to determine the total redshift expression~\eqref{s10} that applies to any spherically symmetric metric in the presence of NED fields. Since astronomers usually measure the kinematic redshift (Doppler effect), we had to determine its general maximum and minimum expressions~\eqref{s14}. Applying these expressions to two NED-metric members of the $h$-family, we have observed amplification of the kinematic redshift when the emitting star moves in the vicinity of the smallest circular geodesic, provided $Q/M \gtrsim 0.79$. This effect allows us not only to distinguish between the NED metrics and the RN BH but also to fix one of the NED-solution parameters if the other two are known a priori, suggesting potential astrophysical relevance.

We have noticed that for moderate and higher values of the electric charge, there are notable discrepancies between the physical and geometric behaviors of the NED metrics and the RN BH. These include the shadow of the BH solution and the kinematic redshift. For the weak deflection angle, however, the results are very similar up to the $1/b^3$ approximation due to the far-field behavior of the NED metrics.

Possible avenues for future work are discussed as follows. In general, nonlinear electromagnetic fields are expected to play a major role in the strong field regime. Therefore, it would be interesting to investigate the deflection angle in the strong field limit~\cite{Bozza:2002zj,Eiroa:2010wm,Tsukamoto:2016jzh}, with the aim of capturing possible effects associated with NED fields. Moreover, throughout this paper, we have considered NED models that depend solely on $F$. For an NED structure depending on the two electromagnetic scalars, we may have birefringence~\cite{Novello:1999pg}. In this case, light rays can follow different paths according to their polarization. This opens up the possibility of investigating the relationship between this phenomenon and the gravitational and kinematic redshifts. Furthermore, our results were restricted to the BH case, but it is known that the NED framework can be used to obtain other compact objects, such as wormholes~\cite{Bronnikov:2017sgg,Bronnikov:2021uta}, which also makes studies in this vein worthwhile. Finally, in real astrophysical scenarios, BHs are expected to spin. Consequently, a further extension of the results presented in this work is to consider rotating geometries sourced by NED models.

%%%%%%%%%%%%%%%%%%%%%%%%%%%%%%
\section*{Data availability statement}
%%%%%%%%%%%%%%%%%%%%%%%%%%%%%%

The data cannot be made publicly available upon publication because no suitable repository exists for hosting data in this field of study. The data that support the findings of this study are available upon reasonable request from the authors.

%%%%%%%%%%%%%%%%%%%%%%%%%%%%%%
\ack
%%%%%%%%%%%%%%%%%%%%%%%%%%%%%%

M. A. A. de Paula is supported by CNPq/PDJ 150589/2025-5.

\section*{ORCID iDs}

Marco A. A. de Paula \href{https://orcid.org/0000-0002-9808-7495}{https://orcid.org/0000-0002-9808-7495} \\ Mustapha Azreg-A\"{\i}nou \href{https://orcid.org/0000-0002-3244-7195}{https://orcid.org/0000-0002-3244-7195} 

\appendix

\section{No-go theorem revisited}\label{appx}

In Ref.~\cite{Bronnikov:2000vy}, a no-go theorem established that for SSS solutions with nonzero electric charge, satisfying the Maxwell weak-field limit, i.e., $\mathcal{L}(F) \to F$ and $\mathcal{L}_F \to 1$ as $F \to 0$, RBHs cannot exist, whether the configuration is purely electric or dyonic. The theorem assumes that $\mathcal{L}(F) \to F$ as $F \to 0$ both at the center and at spatial infinity, thus excluding electrically charged RBHs only under these conditions.

The new solutions presented in Ref.~\cite{Azreg-Ainou:2025tuj} circumvent the theorem by constructing models where $\mathcal{L}(F)$ possesses multiple branches rather than a single-valued form. This structure enables the theory to satisfy the Maxwell limit at infinity without imposing it at the core. At spatial infinity, $\mathcal{L}(F)$ recovers the Maxwell limit ($\mathcal{L} \to F$), ensuring the solutions replicate LED behavior in the far field. Conversely, at the center, where $F$ also approaches zero, $\mathcal{L}(F)$ takes a non-Maxwell form, providing the necessary conditions to regularize the center.

To put it into perspective, let us consider the fully RBHs considered in this paper. At the center, i.e., as $r \to 0$, the limit $F \to 0$ for the $h = 0$ case (see Sec.~\ref{sech0}) leads to
\begin{equation}
	\mathcal{L}(F) = \frac{5184 M^4}{\pi ^3 \left(2+\sqrt{2}\right)^2 Q^6} + \mathcal{O}\left[F^{1/3}\right],
\end{equation}
while for the $h = 2$ case (see Sec.~\ref{sech2}), we obtain
\begin{equation}
	\mathcal{L}(F) = \frac{2048 M^4}{\pi ^4 Q^6}+ \mathcal{O}\left[F^{1/3}\right],
\end{equation}
Therefore, at the center, we notice that as $F \to 0$, $\mathcal{L}(F)$ tends to a constant plus a term proportional to $F^{1/3}$, which is different from the asymptotic behavior at infinity, i.e., $\mathcal{L}(F) \to F$.

In summary, by decoupling the behavior at the center from the asymptotic behavior at infinity, it is possible to generate fully regular electrically charged BH solutions that do not violate the Maxwell-like behavior at the far field. {We also point out that multi-valued Lagrangians can be related to phase transition mechanisms in spacetime, see, e.g., Refs.~\cite{dymnikova2015existence,Gao:2025rep,Burinskii:2002pz}. While a rigorous thermodynamic analysis of these transitions is reserved for future study, since they are beyond the scope of this work, the model studied here uses this mechanism as an effective approach to regularize the central singularity.}

\section{No regular metrics in GR minimally coupled to LED}\label{appx2}

As discussed in Sec.~\ref{sec:background}, it is well known that there are no regular metrics that are solutions of GR minimally coupled to LED. Below is a simple but insightful justification for this. In LED, Eq.~\eqref{3c} yields $F=-2Q^2/r^4$, and the KS reduces to
\begin{equation}
	\label{KSLED}\mathcal{K_{\text{S-LED}}}=\frac{20 Q^4}{r^8}-\frac{24 Q^2 (1-f(r))}{r^6}+\frac{12 (1-f(r))^2}{r^4}\,,
\end{equation}
where $Q$ denotes the electric or magnetic charge in this case. It is clear that there is no way to make $\mathcal{K}_{\text{S-LED}}$ regular at the center $r=0$ whether $f(r)$ is regular there or not. The reader may object and claim that the massless metric function $f(r)=1+\text{const}/r^2$ could remove the divergence at $r=0$. First of all, we must have $\text{const}=Q^2$ but this solves only~\eqref{3b} and not~\eqref{3a}. Moreover, if we assume $\text{const}\neq Q^2$ then we have to choose a complex $\text{const}=\pm \ii (\sqrt{6}\pm 3\ii)Q^2/3$ (where $\ii ^2=-1$) to be able to remove the singularity of $\mathcal{K_{\text{S-LED}}}$ at $r=0$. In other words, it is strictly impossible to regularize $\mathcal{K}_{\text{S-LED}}$ at $r=0$ within LED due to the divergent $r^{-8}$ term.

%%%%%%%%%%%%%%%%%%%%%%%%%%%%%%
\section*{References}
%%%%%%%%%%%%%%%%%%%%%%%%%%%%%%

\bibliographystyle{unsrt}
\bibliography{ref.bib}

\end{document}